\documentclass[useAMS,usenatbib]{mn2e}
\usepackage{amsmath}
\usepackage{graphicx,amssym,color}
\newcommand{\bc}{\begin{center}}
\newcommand{\ec}{\end{center}}

\def\SN{\textsc{S}n}                    
\def\SNII{\textsc{S}n\textsc{ii}}       
\def\SNIa{\textsc{S}n\textsc{i}a}       
\def\AGB{\textsc{AGB}s}                 

\def\Nbins{\textit{n$_{\rm bins}$} }     
\def\Nmet{\textit{n$_{\rm met}$} }       
\def\metar{\textsc{returnedmet} }      

\def\Nmetp{$\textstyle{\textit{n$_{\rm met}$}}+2$ }

\title[Milky Way Galaxies in a Hierarchical Model]
      {Elemental Abundances in Milky Way-like Galaxies from a Hierarchical
        Galaxy Formation Model}   
\author[G.~De Lucia et al.]
       {Gabriella De Lucia$^{1}$\thanks{Email: delucia@oats.inaf.it}, 
        Luca Tornatore$^{1}$, 
        Carlos S. Frenk$^2$, Amina Helmi$^3$, Julio F. \newauthor  Navarro$^4$, 
        Simon D. M. White$^5$\\
        $^1$INAF - Astronomical Observatory of Trieste, via G.B. Tiepolo 11, 
        I-34143 Trieste, Italy\\
        $^2$Institute of Computational Cosmology, University of Durham, 
        Science Laboratories, South Road, Durham DH13LE\\
        $^3$Kapteyn Astronomical Institute, University of Groningen,
        P.O. Box 800, 9700 AV Groningen, the Netherlands\\
        $^4$Dept. of Physics and Astronomy, University of Victoria, 
        Victoria, BC Canada V8P 5C2\\
        $^5$Max Planck Institut f\"ur Astrophysik Karl-Schwarzschild-Str. 1, 
        85741 Garching, Germany}
\begin{document}

\pagerange{\pageref{firstpage}--\pageref{lastpage}} 
\pubyear{2013}

\maketitle

\label{firstpage}

\begin{abstract}
We develop a new method to account for the finite lifetimes of stars and trace 
individual abundances within a semi-analytic model of galaxy formation. At 
variance with previous methods, based on the storage of the (binned) past star 
formation history of model galaxies, our method projects the information about 
the metals produced by each simple stellar population (SSP) in the future. 
Using this approach, an accurate accounting of the timings and properties of 
the individual SSPs composing model galaxies is possible. We analyse the 
dependence of our chemical model on various ingredients, and apply it to 
six simulated haloes of 
roughly Milky Way mass and with no massive close neighbour at z=0. For all 
models considered, the [Fe/H] distributions of the stars in the disc component 
are in good agreement with Milky Way data, while for the spheroid component 
(whose formation we model only through mergers) these are offset low with 
respect to observational measurements for the Milky Way bulge. This is a 
consequence of narrow star formation histories, with relatively low rates of 
star formation. The slow recycling of gas and energy from supernovae in our 
chemical model has important consequences on the predicted star 
formation rates, which are systematically lower than the corresponding rates 
in the same physical model but with an instantaneous recycling approximation. 
The halo that resembles most our Galaxy in terms of its global properties also 
reproduces the observed relation between the average metallicity and luminosity 
of the Milky Way satellites, albeit with a slightly steeper slope.
\end{abstract}


\section{Introduction}
\label{sec:intro}

The distribution and amount of heavy elements in different baryonic components
of the Universe depend strongly on the details of the physical mechanisms that
regulate the evolution of baryons as a function of cosmic time (among recent
studies, see e.g. \citealt*{DeLucia_Kauffmann_White_2004};
\citealt{Oppenheimer_and_Dave_2006}; \citealt{Tornatore_etal_2010}). As a
consequence, chemical abundance studies provide considerable information about
the sequence of events that characterise the formation and evolution of
galaxies and, more generally, on the relative importance of the various
gas-dynamical processes that determine the observed cosmic evolution of
baryons.

Two excellent examples of such an `archaeological' approach applied to galaxy
evolution studies are: inferences on the star formation history of elliptical
galaxies and its dependence on the galaxy stellar mass
(e.g. \citealt*{Worthey_Faber_Gonzalez_1992}; \citealt{Matteucci_1994};
\citealt*{Thomas_Greggio_Bender_1999}; \citealt*{Graves_Faber_Schiavon_2009}),
and chemical evolution studies aimed at constraining the formation history of
different components of our Milky Way (see e.g. the classical review by
\citealt{Tinsley_1980}; \citealt{Matteucci_2008} and references therein). Both
kinds of studies take advantage of the so-called $\alpha$-elements (these
include O, Mg, Si, S, Ca, and Ti), and of their enhancements relative to
Fe. $\alpha$-elements are released mainly by supernovae type II, originating
from massive progenitors with relatively short main sequence life-times 
($\sim$3$-$20~Myr). In contrast, the main contribution to the Fe-peak elements 
is given by supernovae type Ia, that are believed to originate from 
thermonuclear explosions of white dwarf stars whose progenitors have life-times 
ranging from $\sim$30~Myr up to several Gyr. Therefore, the [$\alpha$/Fe] 
abundance ratio can be used as a powerful diagnostic of the star formation 
history and/or of the initial stellar mass function 
\citep{Tinsley_1979,McWilliam_1997}.

During the last decades, accurate abundance measurements have been collected
for a large number of individual stars in our Milky Way and in a number of the
brightest satellites of the Local Group, as well as for thousands of nearby
galaxies. A range of chemical elements have been measured in stars, in the cold
interstellar gas, and in the hot intergalactic medium, back to an epoch when
the Universe was only about one tenth of its current age. More and better data
will come from ongoing and future observational programmes. The interpretation
of this vast amount of data requires sophisticated chemical models, able to
follow self-consistently the evolution of individual element
abundances. 

Unfortunately, most of the detailed chemical models mentioned above are not
embedded in the framework of the currently favoured cosmological model for
structure formation. In contrast, all theoretical models of galaxy evolution in
a cosmological context have traditionally relied on an instantaneous recycling
approximation. These methods include hydrodynamical simulations and
semi-analytic models of galaxy formation. The former introduces an explicit
treatment of gas physics in N-body simulations. In the second approach, the
evolution of the baryonic component is modelled invoking `simple', yet
physically and observationally motivated `prescriptions'. The basic assumption
is that galaxies form when gas condenses at the centre of dark matter
haloes. Star formation, feedback processes, chemical enrichment, etc. take then
place according to analytical laws which are based on theoretical and/or
observational arguments. Adopting this formalism, it is possible to express the
full process of galaxy evolution through a set of differential equations that
describe the variation in mass as a function of time of the different galactic
components (e.g.  stars, gas, metals), and that are coupled to the merger
history of the dark matter haloes extracted from the $N$-body
simulations \citep[for a review, see e.g.][]{Baugh2006}.

The situation has improved significantly in recent years with detailed chemical
schemes being incorporated both in hydro-dynamical
simulations \citep{Tornatore_etal_2004,Scannapieco_etal_2005,Wiersma_etal_2009,Few_etal_2012}
and in semi-analytic models of galaxy
formation \citep{Nagashima_etal_2005,Cora_2006,Pipino_etal_2009,Arrigoni_etal_2010,Benson_2012,Yates_etal_2013}.

In this paper we present a new, fully self-consistent implementation of
chemical enrichment within a semi-analytic model of galaxy formation, and apply
it to N-body simulations of Milky Way-like haloes. The layout of the paper is
as follows: in Section~\ref{sec:simsam}, we provide a brief description of the
simulations used in our study, and summarise the relevant elements of the
reference galaxy formation model. In Section~\ref{sec:chemsam}, we discuss the
main ingredients of our chemical evolution model, and describe in detail how we
have updated our galaxy formation model to relax the instantaneous recycling
approximation. In Section~\ref{sec:ingredients}, we analyse the influence of
the different model ingredients, and in Section~\ref{sec:compareIRA} we discuss
how the basic properties predicted by our model for simulated Milky Way-like 
galaxies are affected by using the updated chemical evolution scheme. In
Section~\ref{sec:results}, we show our basic predictions for the chemical
distributions of stars in the spheroid and disc of our model Milky Way-like 
galaxies, as well as for the mass-luminosity relation of satellite galaxies. 
Finally, we discuss our results and summarise our conclusions in
Section~\ref{sec:discconcl}.

\section{The simulations and the galaxy formation model}
\label{sec:simsam}

In this paper, we make use of the simulations carried out within the
Aquarius Project \citep{Springel_etal_2008}. A sample of six haloes of roughly
Milky Way mass and with no massive close neighbour at $z=0$ were selected from
the Millennium II Simulation \citep{Boylan-Kolchin_etal_2009}. The haloes and
their immediate environment were then re-simulated at higher resolution within 
the original periodic cube of side $100\,h^{-1}\,{\rm Mpc}$ using the following
cosmological parameters: $\Omega_{\rm m}=0.25$, $\Omega_{\Lambda}=0.75$, 
$\sigma_8=0.9$, $n=1$, and Hubble constant 
$H_0 = 100\, h\, {\rm km} \, {\rm s}^{-1}\, {\rm Mpc}^{-1}$ with $h=0.73$. The
adopted value for the normalisation of the power spectrum ($\sigma_8$) is
consistent with WMAP first-year results, but is higher than more recent
estimates. However, as shown in previous work (\citealt{Wang_etal_2008}; see
also \citealt{Guo_etal_2013}), this will not affect the results of our galaxy
formation model significantly.

\begin{table}
\begin{tabular}{lccc}
\hline
Name & $m_{p}$ & $M_{200}$ & $R_{200}$\\
     & ($M_{\odot}$) & ($M_{\odot}$) & (kpc)\\
\hline
\hline
Aq-A-3 & $4.91 \times 10^{4}$ & $1.84 \times 10^{12}$ & 245.64 \\ 
\hline
Aq-A-2 & $1.37 \times 10^{4}$ & $1.84 \times 10^{12}$ & 245.88 \\
Aq-B-2 & $6.45 \times 10^{3}$ & $8.19 \times 10^{11}$ & 187.70 \\
Aq-C-2 & $1.40 \times 10^{4}$ & $1.77 \times 10^{12}$ & 242.82 \\
Aq-D-2 & $1.40 \times 10^{4}$ & $1.77 \times 10^{12}$ & 242.85 \\
Aq-E-2 & $9.59 \times 10^{3}$ & $1.19 \times 10^{12}$ & 212.28 \\
Aq-F-2 & $6.78 \times 10^{3}$ & $1.14 \times 10^{12}$ & 209.21 \\
\hline
\end{tabular}
\caption{Basic numerical parameters of the simulations used in this study. The 
  table lists: the simulation name, the particle mass ($m_{p}$), the virial 
  mass of the halo ($M_{200}$) and the corresponding virial radius ($R_{200}$). 
  For all these simulations, the Plummer-equivalent gravitational softening 
  length has been set to $65.8$~pc. \label{tab:sims}}
\end{table}

The simulated Milky Way haloes have virial masses ($M_{200}$\footnote{In this
  study, $M_{200}$ is defined as the mass contained within the radius enclosing
  a mean density corresponding to 200 times the critical value at the redshift
  of interest.}) in the range $0.8$--$1.8\times10^{12}\,{\rm M}_{\sun}$, 
broadly consistent with observational estimates for the Milky Way
\citep{Battaglia_etal_2005,Smith_etal_2007}. Haloes were simulated at different
levels of resolution with particle mass $m_{p}$ ranging between
$3.143\times10^6\,{\rm M}_{\sun}$ (for Aq-A-5) and $1.712\times10^3\,{\rm
M}_{\sun}$ (for Aq-A-1). In the following, we will focus on the resolution
level 2, which is available for all six Aquarius haloes. To verify the
convergence of our physical model and to analyse the influence of different
ingredients of the chemical model, we will also use a lower level of resolution
for the Aq-A halo. This simulation (Aq-A-3) has been run with a dark matter
particle mass $m_{p}=4.91\times 10^4\,{\rm M}_{\sun}$, and a Plummer equivalent
gravitational softening length of $120.5$~pc. Numerical parameters of all the
simulations used in this work are summarised in Table~\ref{tab:sims}.

Simulation data were stored for 111 (for Aq-F-2), 128 (Aq-B-2 to Aq-E-2), 512
(Aq-A-3), and 1024 (for Aq-A-2) snapshots. Each of these was processed using a
standard friends-of-friends (FOF) algorithm with linking length equal to 0.2
times the mean inter-particle separation, and with the algorithm {\small
  SUBFIND} \citep{Springel_etal_2001} to identify all self-bound dark matter
substructures in each FOF halo. Merger trees were then constructed linking
each subhalo to a unique descendant in the subsequent snapshot (considering one
every eight snapshots for Aq-A-2 and Aq-A-3, and all snapshots for the other 
Aq haloes) using software developed for the Millennium I simulation (for 
details, see \citealt{Springel_etal_2005} and 
\citealt{DeLucia_and_Blaizot_2007}). A study based on the Millennium II by
\citet{Boylan-Kolchin_etal_2010} has shown that the Aquarius haloes span the
diverse properties of Milky Way haloes, in terms of assembly history and
present day structure. Haloes A and C are found to have formed somewhat earlier
than average. Halo F is somewhat unusual since it experienced a recent major
merger, but it is found to be quite typical in most other properties.

The merger trees described above are used as input for a galaxy formation
model. The specific model used in this work is the `ejection model' described
by \citet*{Li_DeLucia_Helmi_2010}. This is based on the model described in
detail by \citet[][and references therein]{DeLucia_and_Blaizot_2007}, but has
been modified to follow more accurately processes on the scale of the Milky
Way's satellites (for details, we refer to \citealt{DeLucia_and_Helmi_2008}
and \citealt{Li_DeLucia_Helmi_2010}). The same model has already been combined
with the Aquarius dark matter simulations by \citet{Starkenburg_etal_2013} to
investigate the properties of the satellites of Milky Way-like galaxies.

In our galaxy formation model, baryons are in four different components: stars
in galaxies, cold gas in galaxy discs, hot gas associated with FOF haloes, and
ejected gas that has been reheated and expelled outside haloes by
supernova-driven winds. This ejected component can later be reincorporated
into the hot gas component that is associated with the central galaxy of the 
FOF. 

The circulation of baryons between different components is modelled using
simplified yet physically and/or observationally motivated prescriptions. The
chemical evolution model used in previous work tracked metal production in the
instantaneous recycling approximation with a constant yield ($Y=0.03\,{\rm
M}_{\sun}$) of heavy elements for each solar mass of new stars, and a constant
fraction ($R=0.43$ for a Chabrier IMF) of stellar mass returned immediately to
the interstellar medium. All material lost from stars was assumed to be
perfectly mixed with the cold gas component, except in low-mass haloes (${\rm
M}_{\rm vir} < 3\times 10^{10}\,{\rm M}_{\sun}$) where it was assumed to be
ejected directly into the hot gas component. Metals were exchanged between
different baryonic phases according to specific feedback and reincorporation
schemes. Generally, flows of metals were proportional to the mass flows between
different baryonic components. For a detailed description of the scheme adopted
to model chemical enrichment, we refer to
\citet*{DeLucia_Kauffmann_White_2004}. In the next section, we describe how 
this model has been updated in order to account for the finite life-time of 
stars and delayed gas recycling and chemical enrichment.
  
As for most published semi-analytic models, the morphological information for
our model galaxies is given by the relative contribution of the `disc' and the
`bulge'. The stellar disc component originates from quiescent star formation
taking place in the gaseous disc forming following cooling of gas within the
parent dark matter halo. In our previous work, the bulge component was assumed
to form through both mergers and disc instability
(see \citealt{DeLucia_and_Helmi_2008} for details). An explicit modelling of
the disc instability would be required to compare our model predictions with
observational measurements of the `bulge' of the Milky Way. The latter has a
complex 3D structure that suggests that significant disc and/or bar dynamical
evolution has taken place \citep[e.g.][and references
therein]{Dwek_etal_1995,Stanek_etal_1994,Vasquez_etal_2013}. Unfortunately,
however, the adopted modelling of disc instability is rather uncertain (see
e.g. \citealt{DeLucia_etal_2011} and references therein). In the following, we
have decided to switch off this channel for (pseudo)bulge formation. Therefore,
our `bulge' (spheroid) component should correspond to part of the Milky Way
bulge and of its inner stellar halo. In addition, we do not model here the
formation of the stellar halo from tidal stripping of satellites galaxies (as
done for example in \citealt{DeLucia_and_Helmi_2008}
and \citealt{Cooper_etal_2010} - this will be the subject of a future work),
and do not distinguish between a thin and a thick disc component. Since the
formation of the latter might be related to the formation of the stellar halo,
and given it represents a small fraction (about $10$--$20$ per cent) of the
stellar mass in the disc, only comparisons between our model predictions and
the thin disc are appropriate.

\section{The chemical evolution model}
\label{sec:chemsam}

During their lifetime, stars lose mass via stellar winds. Some stars ultimately
undergo supernova (\SN) explosions polluting the interstellar medium with both
newly produced and earlier synthesised metals. The rates and timescales of both
these phenomena depend on the mass of the stars. In our previous work, we
neglected the delay between star formation and the recycling of gas and metals.
Our reference model was therefore not able to take into account the evolution
of individual element abundances, and, in particular, did not describe well the
production of elements around the iron-peak, which are mainly produced by
\SNIa~with delay times ranging from a few tens of million years to several 
billion years.  In this section, we describe how our earlier model has been 
updated to relax the instantaneous recycling approximation (IRA), and provide a 
more detailed and more realistic treatment of chemical evolution.

\subsection{Basic equations and model ingredients}

The rate at which enriched gas is lost by stars and supernovae into the
interstellar medium of a galaxy depends on the integral past star formation
history of all its progenitors. At any given time $t$, the rate at which the
element $i$ is injected into the interstellar medium can be calculated using
the following equation:

\begin{eqnarray}
\label{eq:chemeq}
 R_{Z_i}(t) &=& R_{\rm Ia}(t) \cdot Q_{\rm i}^{\SNIa}\nonumber \\
&+& \int_{M_{\rm L}}^{M_{\rm SnII}} \Psi(t-\tau_m) Q_{i}^{\AGB}(m,Z) \phi(m) dm\\
&+& \int_{M_{\rm SnII}}^{M_{\rm U}} \Psi(t-\tau_m) Q_{i}^{\SNII}(m,Z) \phi(m) dm 
\nonumber
\end{eqnarray}

In this work, we assume ${\rm M}_L=0.1\,{\rm M}_{\sun}$, ${\rm M}_U=100\,{\rm
M}_{\sun}$ and ${\rm M}_{\rm SnII} = 8\,{\rm M}_{\sun}$.  The various terms in
Eq.~\ref{eq:chemeq} represent the contribution of stars in different mass
ranges. In particular:
\begin{description}
\item[(i)] the first term represents the contribution from \SNIa, and is the
  product between the rate of \SNIa~at time $t$ ($R_{\rm Ia}$ - see
  sec.~\ref{sec:snia} for details), and the yields from \SNIa~($Q_{\rm
  i}^{\SNIa}$ - in our case, these do not depend on the metallicity or mass of
  the progenitors);
\item[(ii)] the second term represents the contribution of low mass stars. 
  Note that this integral extends to masses above the minimum mass of \SNIa~
  producing binary systems ($\sim$3~${\rm M}_{\sun}$). Therefore, this term
  explicitly accounts for the ejection of metals during the AGB phase of these
  stars, prior to the supernova explosions;
\item[(iii)] the last integral represents the contribution from stars with mass
  larger than ${\rm M}_{\rm SnII}$, that end their life as \SNII.
\end{description}

$\Psi(t)$ is the star formation rate at time $t$, $\phi(m)$ is the initial mass
function, and $\tau_m$ is the lifetime of a star of mass $m$. The quantities
$Q_i^{\AGB}(m,Z)$ and $Q_i^{\SNII}(m,Z)$ represent the yields array for the 
element $i$ from \AGB~and \SNII, and depend on the mass and the total 
metallicity of the star. In principle, our model can take into account the 
dependence on the detailed chemical composition of the stars so that the 
factors $\Psi(t-\tau_{\rm m}) Q_i(m,Z)$ in the above equation should read as:

\begin{equation}
\int \Psi(t-\tau_{\rm M}, Z_{1,\dots,{\rm N}}) Q_i(m, Z_{1,\dots,{\rm N}})
dZ_{1,\dots,{\rm N}}
\end{equation}

\noindent
Since, however, we trace an average star formation event for each time-step, and
the yields tables we employ depend only on the total metallicity, this double
integral is not used in this work. 

\subsubsection{Supernovae Type Ia}
\label{sec:snia}

To compute the first term of Eq.~\ref{eq:chemeq}, it is necessary to know the
rate of \SNIa, which depends on the nature of the progenitor
systems. Unfortunately, this is still poorly understood. The widely accepted
scenario is that of a carbon and oxygen white dwarf (WD) that accretes mass 
from a companion star in a binary system. As it approaches the Chandrasekhar
limit ($1.4\,{\rm M}_{\sun}$), the highly degenerate WD ignites thermonuclear
fusion in an explosive event that completely destroys the star 
\citep[see e.g.][]{Wang_and_Han_2012,Hillebrandt_etal_2013}. One of the key 
unknowns is the nature of the companion star: another WD in the `double 
degenerate model' \citep{Iben_and_Tutukov_1984,Webbink_1984}, a main sequence 
dwarf or an evolved giant in the `single degenerate model' 
\citep{Whelan_and_Iben_1973}. Once the progenitor model is chosen, the 
\SNIa~rate can be computed by convolving the star formation rate with the 
distribution of the explosion times, that is usually referred to as `delay 
time distribution' (DTD) function. 

Recently, 
\citet[][see also \citealt{Ruiz-Lapuente_and_Canal_1998}]{Greggio_2005} has 
introduced a relatively simple and general
analytical formulation for the \SNIa~rate that does not depend on the details
of the progenitor model, and allows an efficient investigation of the influence
of different DTDs on the chemical properties of interest. In this formalism,
the rate of SNIa at an epoch $t$ can be written as:
\begin{equation}
\label{eq:Ia}
R_{Ia}(t) = k_\alpha\, \int_{\tau_i}^{{\rm min}(t, \tau_x)}\;
A(t-\tau)\Psi(t-\tau)DTD(\tau){\rm d}\tau 
\end{equation}
where $A(t-\tau)$ represents the realization probability of a given
\SNIa~scenario from the stellar population born at the time $(t-\tau)$,
$\tau_i$ is the minimum delay time and corresponds to the minimum evolutionary
lifetime of the \SNIa~precursors, and $\tau_x$ is the maximum delay time. We
assume $\tau_i = 29$~Myr (that is, in our formulation, the lifetime of a star
with mass $8\,{\rm M}_{\sun}$) and $\tau_x = 20$~Gyr (the lifetime of a star
with mass $0.8\,{\rm M}_{\sun}$). In chemical evolution models, $A$ is usually
treated as a free parameter and is tuned to reproduce the present
time \SNIa~rate. In our study, we will also treat A as a free parameter, but
will tune its value by considering the [Fe/H] and [O/Fe] distributions of disc
stars for our model Milky Way galaxies.

The DTD is defined in the range $[\tau_i,\tau_x]$, and is normalised as:
\begin{equation}
\int_{\tau_i}^{\tau_x}DTD(\tau){\rm d}\tau = 1;
\end{equation}
$k_\alpha$ is the number of stars per unit mass in a stellar generation and is 
derived from the initial stellar mass function:
\begin{equation}
k_\alpha = \int_{{\rm M}_L}^{{\rm M}_U}\phi(m){\rm d}m
\end{equation}

Therefore, Eq.~\ref{eq:Ia} allows us to include consistently the \SNIa~events 
in our galaxy formation model by specifying the parameter A and the DTD, and 
ignoring modelling details of the binary population of \SNIa~progenitors. 

Formally, stars that end their life as \SNIa~should not be counted in the
second and third terms of Eq.~\ref{eq:chemeq}. However, in the adopted
formalism for \SNIa, the information on the mass distribution of these stars is
hidden in the parameter $A$, and should be computed explicitly for each
specific model corresponding to the assumed DTD. It is, in general, not
possible to relate the A parameters computed for different scenarios, as done
for example by \citet{Arrigoni_etal_2010}.  Since the expected values for $A$
are typically small, below a few percents, the correction would not influence
significantly the chemical enrichment calculations. Therefore, in the
following, we do not to correct the second and third terms in
Eq.~\ref{eq:chemeq}.

\subsubsection{The distribution function of the delay times for supernovae Ia}
\label{sec:snia_DTD}

As explained in the previous section, different progenitor scenarios
for \SNIa~ predict different distribution functions for their delay
times. Different functions have been proposed in the literature, and
below we will analyse the influence of five different DTDs:

\begin{figure}
\centering
\resizebox{8.7cm}{!}{\includegraphics{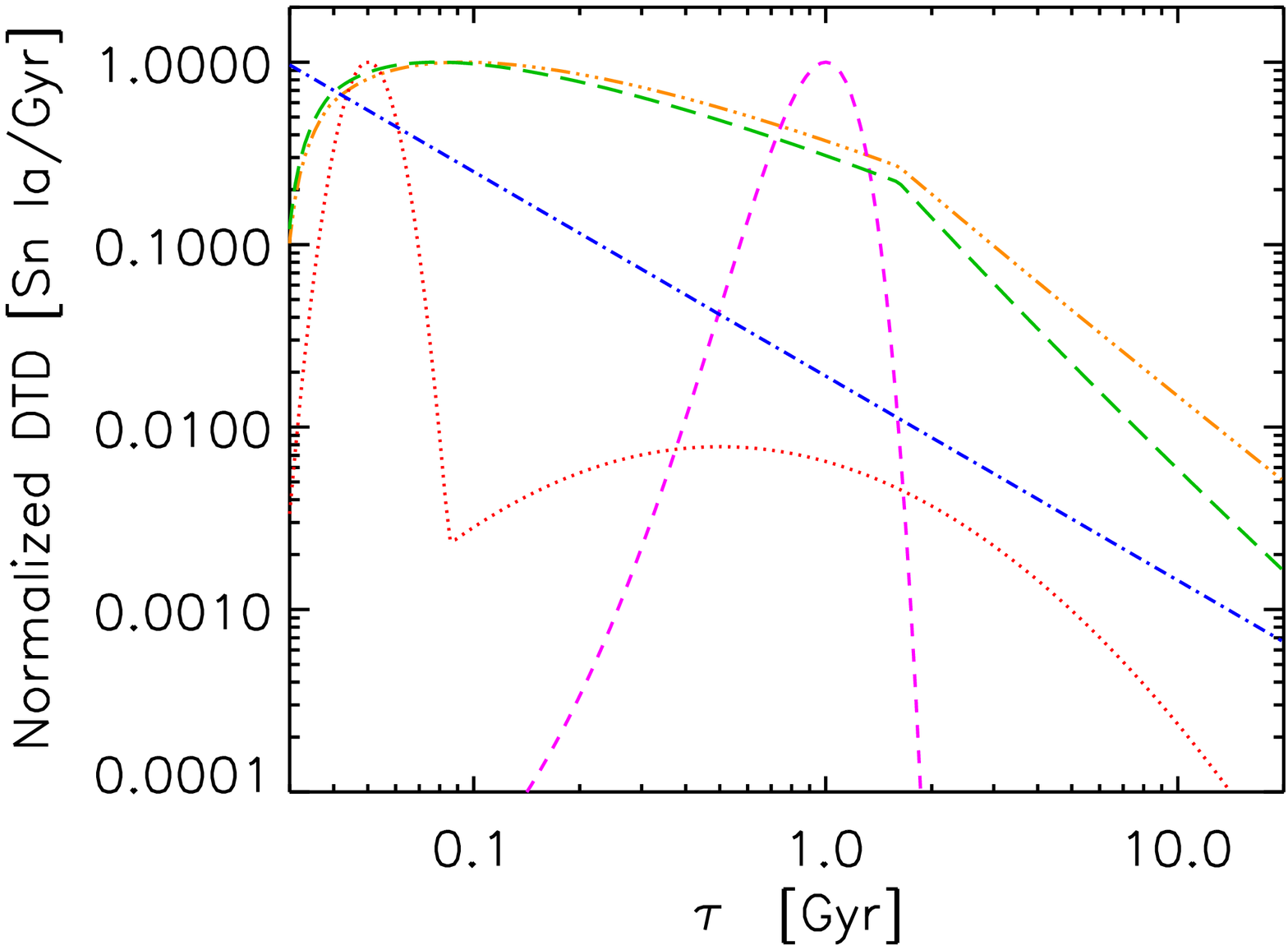}} 
\caption{The normalised DTDs considered in this study (see text for
  details). The blue dot-dashed line corresponds to a power-law DTD with
  slope $-1.12$, as proposed by \citet{Maoz_Mannucci_Brandt_2012}; the magenta
  dashed line to the narrow Gaussian DTD proposed
  by \citet{Strolger_etal_2004}; the red dotted line to the bi-modal DTD
  proposed by \citet{Mannucci_DellaValle_Panagia_2006}; the green long
  dashed and orange dot-dot-dashed lines to two different variations of
  the DTD corresponding to a single degenerate scenario as described
  in \citet{Matteucci_and_Recchi_2001}
  and \citet{Bonaparte_etal_2013}.\label{fig:dtds}}
\end{figure}

\begin{enumerate}

\item a power-law DTD with slope $-1.12$, as proposed by
  \citet*{Maoz_Mannucci_Brandt_2012}. This is shown as a blue dot-dashed
  line in Fig.~\ref{fig:dtds}, and has been inferred from a sample of 132
  supernovae Ia discovered by the Sloan Digital Sky Survey II in about 66,000
  galaxies, with star formation histories reconstructed from the observed
  spectra. This DTD is consistent with the $\sim$t$^{-1}$ form that is
  generally expected in the double degenerate scenario.

\item The narrow Gaussian DTD function from \citet{Strolger_etal_2004}. This
  has been measured from deep Hubble Space Telescope observations of 42
  supernovae Ia in the redshift range $0.2 < z < 1.6$. The analysis by Strolger
  et al. shows that delay time models that require a large fraction of `prompt'
  supernovae (in the work by Strolger et al., all \SNIa~ explosions with delay
  times less than $\sim$2~Gyr) poorly reproduce the observed redshift
  distribution, and are rejected at high confidence level. The best fit to the
  observed data is obtained for mean delay times in the range $2$--$4$~Gyr. The
  corresponding DTD is shown as a magenta dashed line in
  Fig.~\ref{fig:dtds}.

\item The bi-modal DTD proposed by \citet*{Mannucci_DellaValle_Panagia_2006}.
  This has been shown to simultaneously reproduce the \SNIa~rate evolution as a
  function of redshift, the dependence of the \SNIa~rate on host galaxy
  colours, and the increase of \SNIa~rate in radio-loud ellipticals. The
  assumption of this particular DTD implies that 30 to 50 per cent of the
  supernovae Ia explode within $10^8$~yr from the star formation episode (the
  `prompt' component), while the rest have delay times up to several Gyr (the
  `tardy' component). To implement this particular DTD in our model, we use the
  analytic approximation presented in \citet{Matteucci_etal_2006}. This is
  shown as a red dotted line in Fig.~\ref{fig:dtds} and corresponds to a
  prompt component of $\sim$49 per cent.

\item Finally, two slightly different variations of the DTD corresponding to
  the single degenerate scenario, as described
  in \citet{Matteucci_and_Recchi_2001} and \citet{Bonaparte_etal_2013}. The
  corresponding DTDs are shown by the green long-dashed and orange 
  dot-dot dashed lines in Fig.~\ref{fig:dtds} (we will refer to these as `SD
  narrow' and `SD broad', respectively). In these particular models, the
  fraction of supernovae exploding within $10^8$yr is of the order of $\sim$5
  per cent, while the corresponding fraction within $4\times10^8$~yr is
  $\sim$23 per cent for the SD narrow DTD and $\sim$28 per cent for the SD
  broad DTD.
\end{enumerate}

\subsubsection{Initial Mass Function and Stellar Lifetimes}

In order to solve Eq.~\ref{eq:chemeq}, one needs to define two additional
quantities: the initial stellar mass function (IMF) that sets how newborn stars
are distributed as a function of their stellar mass, and the typical lifetime
for a star of given mass. In this work, we adopt a Chabrier IMF 
\citep{Chabrier03}, that is also consistently used to compute the luminosities
of our model galaxies. 

For the stellar lifetimes, we assume:
\begin{displaymath}
\tau(m) = 1.2 \times m^{-1.85} + 0.003 
\end{displaymath}
for stars with mass $>6.6\,{\rm M}_{\sun}$, and:
\begin{displaymath}
\tau(m) = 10^{\left[ 1.338 - \sqrt{1.790 -0.2232\times(7.764-{\rm
        log}(m))}\right]/0.1116\,-\,9} \,{\rm Gyr} 
\end{displaymath}
for stars with mass $\le 6.6\,{\rm M}_{\sun}$ \citep[see][and references
  therein]{Padovani_and_Matteucci_1993}. We have also tested the different
  formulation presented in \citet{Maeder_and_Meynet_1989} and found that this
  ingredient of the chemical evolution model does not influence significantly
  the results discussed in this paper. As discussed
  in \citet{Romano_etal_2005}, different stellar lifetime prescriptions differ
  mostly in the low stellar mass range. Therefore, models adopting different
  prescriptions for the stellar lifetimes differ mostly in the predicted
  evolution for species originating mainly from low-mass stars like $^3$He and
  $^7$Li. We have neglected here the metallicity dependence of stellar
  lifetimes that is, however, not very strong
\citep[][see also Figure 1 in \citealt{Yates_etal_2013}]{Portinari_Chiosi_Bressan_1998}.

\subsubsection{Chemical yields}

Stellar yields define the amount of material that is returned to the
interstellar medium in the form of newly produced elements during the entire
stellar lifetime, and represent one of the most important ingredients of any
chemical evolution model. Different sets of yields are currently available,
from state-of-the-art stellar evolution and nucleosynthesis calculations. 
Unfortunately, given the large uncertainties still affecting these studies, the 
available yields differ significantly from each other. This implies that
chemical evolution models assuming different sets of yields can provide
significantly different results, even when the same star formation history and
initial mass function are assumed.

In a recent study, \citet{Romano_etal_2010} have tested different stellar
nucleosynthesis prescriptions in the framework of a chemical evolution model of
the Galaxy. They show that large uncertainties are still found, in particular
for the majority of the iron-peak elements but also for more abundant species
like carbon and nitrogen. Uncertainties originate from e.g. neglecting stellar
rotation, different rates of stellar mass loss, different treatment for
convection, etc.

Our reference chemical model uses the following set of yields for stars in
different mass ranges:

\begin{enumerate}
  \item \citet{Karakas_2010} for low- and intermediate-mass stars; 
  \item \citet{Thielemann_etal_2003} for supernovae Ia;
  \item \citet{Chieffi_and_Limongi_2004} for supernovae II.
\end{enumerate}

Note that the yields by \citet{Chieffi_and_Limongi_2004} are defined for
stellar masses between 13 and 35~${\rm M}_{\sun}$. For stellar masses between 8
and 13~${\rm M}_{\sun}$, we scale the yields proportionally to the stellar
mass, while for more massive stars we use the values corresponding to the
highest stellar mass tabulated. Below, we also test the influence of
alternative sets of yields. In particular, we use the yields
by \citet{vandenHoek_and_Groenewegen_1997} for low and intermediate-mass stars,
and those from
\citet[][their case B]{Woosley_and_Weaver_1995} for supernovae II
\footnote{When testing the yields by \citet{Woosley_and_Weaver_1995}, we 
accounted for the contribution of radioactive $^{56}$Ni to Fe in stars with
mass larger than $12\,{\rm M}_{\sun}$.}. New sets of stellar yields have been
recently calculated for low to massive stars using the same nuclear reaction
network \citep{Pignatari_etal_2013}. It will be worth testing the influence of
these new yields in the context of galaxy formation models like the one
presented in this study.

\subsection{Implementing chemical evolution in a galaxy formation model}

Implementing Eq.~\ref{eq:chemeq} in the framework of our galaxy formation model
is, in principle, straightforward. Each star formation episode could be treated
as a simple stellar population (SSP) so that an accurate treatment of chemical
evolution would simply require to store the information relevant to each star
formation event, along with any parameter that might be useful to describe the
future evolution of the SSP (e.g. its chemical composition or some proxy of it,
the IMF in case it is assumed to depend on some physical condition of the gas,
etc.). In practice, however, this approach is unfeasible because of the very 
large memory requirement implied. 

\subsubsection{Storing the past average star formation history}

One obvious possible approach to reduce the memory load is to re-bin the past
star formation history of each galaxy (and eventually of its different
components) in a number of steps smaller than the actual number of time-steps
used in the galaxy formation model. One can then use the stored information to
compute the metal restitution rates at any time during model integration. This
is the approach that has been adopted traditionally in semi-analytic models
that include a detailed chemical evolution model
\citep{Nagashima_etal_2005,Arrigoni_etal_2010,Benson_2012,Yates_etal_2013}.

When accounting for the contributions from both short--living (i.e. \SNII) and
long--living stars (i.e. \SNIa~and asymptotic giant branch stars -- \AGB), the
time interval of interest is comparable to the lifetime of the Universe. Since,
however, most of the metals are produced by \SNII~and by the first \SNIa, it is
convenient to use a finer time resolution close to each star formation event.
In practice, each galaxy is associated with an array carrying the information
relative to the mass of stars formed and the metallicity of the cold gas at the
time of the star formation event. These quantities might not be stored in a
linear time grid: one can use, for example, an `age grid' where the first bin
represents $t=t_{\rm now}$ and the size of the bins increases as one moves to
older ages (see e.g. Appendix A of \citealt{Arrigoni_etal_2010} and Sec.~5.1 of
\citealt{Yates_etal_2013}; in the latter study, the time-bins are chosen to 
coincide with the internal time-steps of the galaxy formation code, and are set 
by the available snapshots of the background simulation).

As the galaxy evolves in time, the information stored in closer (and finer
bins) need to be pushed to older ages, so that each entry of the array stores
the average galaxy star formation rate in the time interval of the past it 
refers to. All the parameters needed to describe the evolution of the SSP will 
also be averaged so that each entry of the star formation array stores an {\it
  effective} SSP. Although it is reasonable to assume that star formation
events close in time will not be significantly different in terms of the
corresponding physical properties (e.g. metallicity), as soon as the time
difference between the star formation events gets larger, the averaging will
involve increasingly physically different SSPs. In addition, the stars ending
up in a model galaxy, will generally form in several progenitors. By 
summing up the star formation histories corresponding to each progenitor, one 
will necessarily mix populations of intrinsically different metallicity. Since 
the calculation of the  metal production depends on the age and metallicity of 
the SSPs, these averaging processes will mostly affect the contribution from 
long-living stars, whose accounting was the main reason to implement a non-IRA 
algorithm in the first place.

\subsubsection{Projecting metal restitution rates into the future}
\label{sec:future}

\begin{figure*}
\centering
\includegraphics[width=1.0\textwidth]{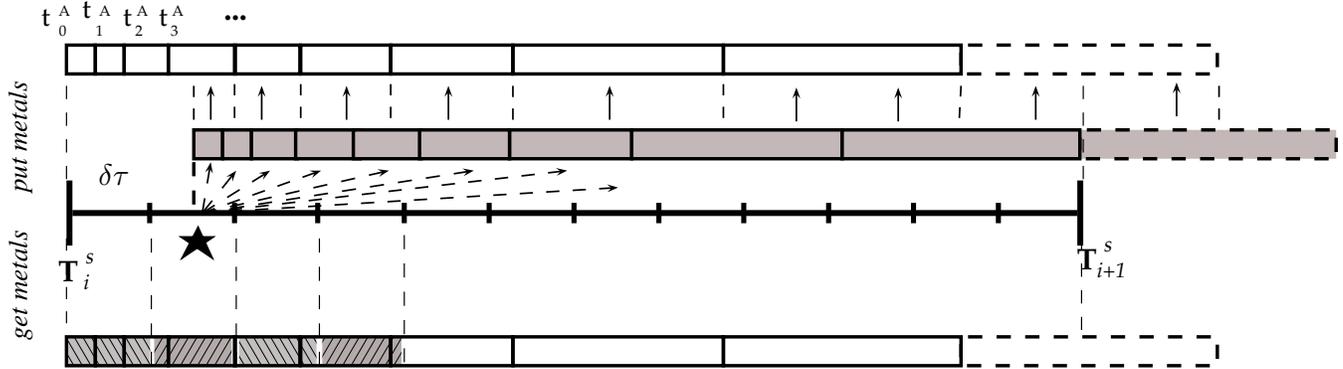}
\caption{Schematic illustration of the method adopted to store the 
  contributions from different types of stars in the future, and incorporate
  the metals in the baryonic gaseous phase of model galaxies during their
  evolution. The thick line shows the time interval between two subsequent
  snapshots. The two arrays at the top and at the bottom of the figure
  represent a `metal restitution array' (\metar) that is associated with each
  model galaxy and contains the mass of elements returned, at any time in the
  future, by the SSPs that constitute the model galaxy under consideration.  At
  each time-step, the code computes the elements produced and adds them to the
  future bins (in case there is an episode of star formation), and then reads
  from the array \metar the amount of metals that needs to be
  re-incorporated. The grey array shown in the figure is a `virtual array' used
  to project metals in the appropriate bins.  \label{fig:project_get_metals}}
\end{figure*}

\begin{figure*}
\centering
\includegraphics[width=1.0\textwidth]{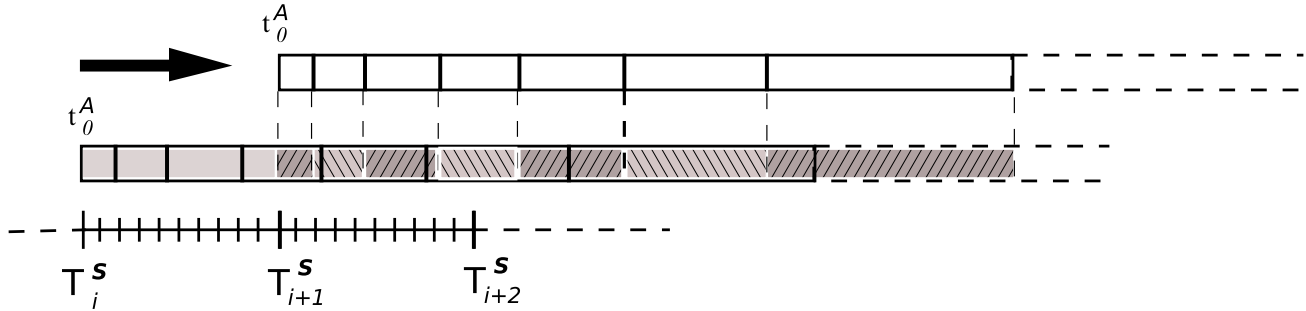}
\caption{At the beginning of each snapshot, the \metar array is pushed forward
  in time and re-aligned with the internal time evolution of the galaxy
  formation model. See text for details.\label{fig:shift_metar}}
\end{figure*}

An alternative approach, that we develop in this study, is to compute the metal
ejection rates every time new stars are formed, and store the corresponding 
information {\it in the future}. As time proceeds, one can evaluate when/if the
appropriate delay times have elapsed and, in that case, incorporate the stored 
metals in the baryonic phases of the model galaxies. Using this approach, an 
accurate accounting of the timings and properties of the SSPs is possible.  

In practice, each model galaxy is associated with a `metal-restitution array'
that contains the mass of every element that will be returned, at some time in
the future, by the SSPs that constitute the galaxy under consideration. The
array will move forward in time as the galaxy evolves, so that also in this
case it is necessary to perform some re-binning. The approximation we make is
that the ejected metals are distributed uniformly in the corresponding time
bins in the future. The total amount of metals formed is, however, computed
using the exact times when new stars are formed, and the correct properties of
the inter-stellar medium at the time of star formation (not averaged, as in the
method illustrated in the previous section). In the following, we detail our
practical implementation of the method summarised above.

To follow the abundance evolution of \Nmet elements, we attach to each galaxy
an array (that we will call \metar in the following) of \Nbins bins, spanning
the time range from $t=t_{\rm now}=0$ to $14$~Gyr in the future. Each time
entry of \metar is an array itself with room for \Nmetp elements. The two extra
elements are used to store the amount of returned $\textrm{H}+\textrm{He}$, and
the energy released by both \SNII~and \SNIa~explosions. The latter is used
self-consistently to model the feedback from supernovae (see below).

Our default option for the time-bins is that of using bins varying in such
a way that each contains the same number of \SN ~events. Our reference
runs adopt a total of 30 bins to follow the chemical evolution. As we will
discuss below, we have verified that our results converge increasing the number
of time-bins. We have also verified that our results are stable against
alternative options for time-bins (e.g log-spaced or just fixed `by hand').

The differential equations governing the evolution of each galaxy are solved
dividing the time interval between two subsequent snapshots ($\Delta T^S =
T^S_{i+1} - T^S_i$) in \textsc{nsteps}$=20$. Let us indicate with
$\delta\tau$ this internal time-step of galaxy evolution. The approach we adopt
is illustrated in Figure~\ref{fig:project_get_metals}. The thick line shows the
time interval between two subsequent snapshots and, in the example illustrated,
a star formation episode occurs during the second integration time-step. The
initial time of the array \metar coincides with the initial time of
integration $T^S_i$, and has a binning that is independent of that used to
model galaxy evolution.

When the simulation starts, the time bins of the chemical evolution are set.
We then build look-up tables that, for each time bin, contain the amount of
every tracked chemical element produced by a SSP of $1\,{\rm M}_{\sun}$. Look-up
tables are calculated for all metallicity bins included in the yield tables
adopted.  In this way, we can significantly speed-up the calculations, that
would otherwise require the estimation of double integrals (see
Eq.~\ref{eq:chemeq}).

Looking at the example shown in Fig.~\ref{fig:project_get_metals}, our method
can be summarized as follows:

\begin{itemize}

\item When the galaxy is evolved from $T^S_i$ to $T^S_i + \delta\tau$, no
star formation takes place. If the \metar is empty (as e.g. at the very
beginning of the simulation), nothing happens and the code just moves to the
following time-step. Otherwise, the code reads from the array \metar the amount
of metals that needs to be re-incorporated (these metals were ejected by
previous generations of stars) and proceeds to the next time-step.

\item During the second time-step, a star formation episode takes place. 
This is treated as a SSP, and the amounts of metals, energy and gas returned at
any time in the future are calculated and stored in the \metar array. The
results of the calculation for the $i$th element and for the $j$th bin of the
array, can be written as:

\begin{equation}
\label{eq:chem_out}
\textstyle{\textsc{out}_i^j} = \int_{T^A_j}^{T^A_{j+1}}
\textstyle{\textsc{chemical evolution}}_i\textstyle{(t, Z)}\,dt
\end{equation}
where $T^A_0,\, T^A_1,\, \dots, T^A_{\Nbins - 1}$ are the times 
corresponding to the beginning of each bin in the \metar array, and
\textsc{chemical evolution} includes all the ingredients described above. 
The subscript $i$ runs from 1 to \Nmetp and refers to a heavy element (if $1\le
i \le \Nmet$), the amount of $\textrm{H}+\textrm{He}$ (if $i
= \textit{n$_{met}$}+1$), or the energy released by supernovae (if $i
= \textit{n$_{met}$}+2$). In our specific model, the evolution depends only on
time and metallicity (of the inter-stellar medium), but more generally it might
depend on other physical properties of the interstellar medium, on the adopted
initial mass function, etc. These additional ingredients could be easily
included in the chemical evolution scheme implemented.

\item In addition to re-incorporating the metals ejected by previous 
generations of stars, the code also adds now into the future bins the elements
produced by the new stars being formed. During the evolution between the two
snapshots, the array \metar is not pushed forward: the grey array shown in
Figure~\ref{fig:project_get_metals} is just a `virtual' array used to
distribute the metals in the appropriate time bin in the future.

\item Only when the code reaches the first time-step of the following 
snapshot, the array \metar is moved forward in time and re-aligned with the
internal time evolution of the galaxy formation model, as illustrated in
Figure~\ref{fig:shift_metar}. The reason for this is practical (to save
memory), and is explained in detail in Appendix A.

\end{itemize}

In practice, our galaxy formation model uses integration time-steps that might
not be negligible with respect to the relevant time-scales of the chemical
evolution. In particular, this applies to elements produced after short delay
times, and at earlier cosmic times, when the time spacing between subsequent
snapshots is larger. Therefore, Eq.~\ref{eq:chem_out} should ideally be
convolved with a non delta--like star formation rate $\Psi$. For simplicity, we
consider all the star-formation events as impulsive in this work.

\section{Influence of chemical model ingredients}
\label{sec:ingredients}

As explained in the previous section, different approximations are made to
include a detailed chemical enrichment treatment in our galaxy formation model,
and different choices are possible for the model ingredients. In this section,
we discuss the effects of using different (i) DTDs, (ii) yields, and (iii)
number of time-bins for the chemical evolution. For this analysis, we will use
the run Aq-A-3 but, as we show below, our physical model converges over the
range of resolution sampled by the simulations considered in this study.

As mentioned earlier, we have fixed the value of the parameter A so as to best
reproduce the [Fe/H] and [O/Fe] distributions of disc stars for the model Milky
Way (in the run Aq-A-3). A value of A$=0.0014$ was chosen for all DTDs adopted
in this study. We note that this value is in good agreement with those adopted
in \citet{Matteucci_etal_2009} and \citet{Yates_etal_2013} (this latter study
uses a different definition for the parameter A - once this is accounted for,
our preferred values are consistent). A single value for A was also found to be
suitable for different DTDs in the studies mentioned above.

\subsection{DTDs}

\begin{figure*}
\bc
\resizebox{18.cm}{!}{\includegraphics{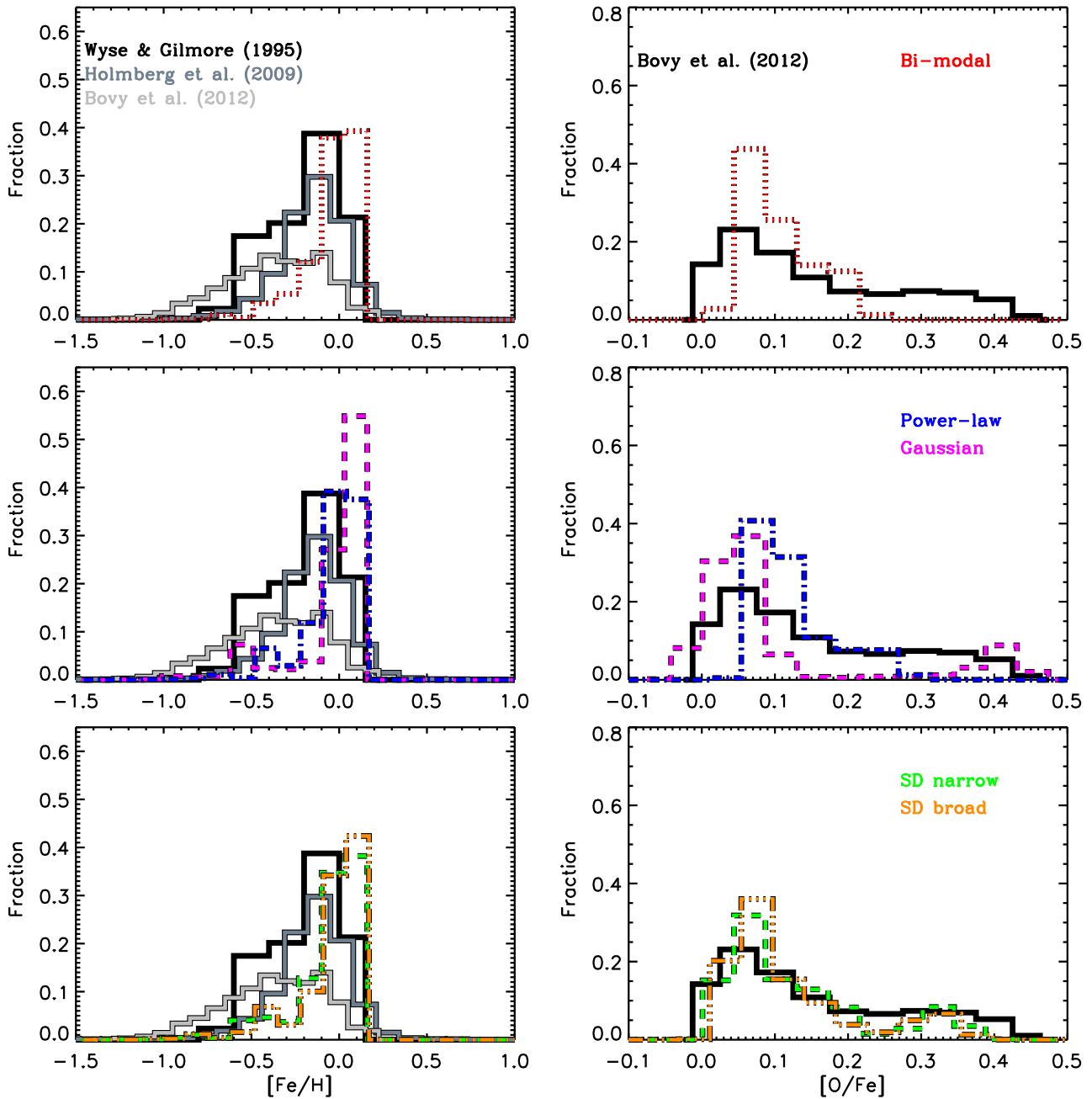}} 
\caption{[Fe/H] (left panels) and [O/Fe] (right panels) distributions for the
  stars in the stellar disc of the run Aq-A-3 (the lowest resolution run used
  in this study), and different assumptions about the DTD (lines of different
  style and colour). Black and grey histograms show observational
  measurements
  by \citet{Wyse_and_Gilmore_1995}, \citet{Holmberg_Nordstrom_Andersen_2009},
  and \citet{Bovy_etal_2012}.}
\label{fig:diskmetdistrDTDs}
\ec
\end{figure*}

Fig.~\ref{fig:diskmetdistrDTDs} shows the [Fe/H] (left panels) and [O/Fe]
(right panels) distributions\footnote{To build these distributions, we attach
to each model galaxy a $20\times20$ matrix that contains the information on the
elemental abundances, and is updated each time an episode of star formation
occurs.} for the stars in the model stellar disc of the simulation Aq-A-3, for
different assumptions about the DTD. Black and grey histograms show
observational measurements, while the coloured lines of different style
show model results (the colour and style coding is the same as in
Fig.~\ref{fig:dtds}). The model [Fe/H] distributions in the left panels are
compared with observational data by
\citet{Wyse_and_Gilmore_1995}, \citet*{Holmberg_Nordstrom_Andersen_2009}, and
\citet{Bovy_etal_2012}. The sample of \citet{Wyse_and_Gilmore_1995} represents
a volume limited sample of long-lived thin disc G stars, while that of
\citet{Holmberg_Nordstrom_Andersen_2009} is based on the Geneva-Copenhagen
Survey of the Solar Neighbourhood, a large ($\sim$16,000) number of F and G
dwarfs with improved astrometric distances from Hipparcos parallaxes. In both
studies, metallicities were derived using a photometric calibration based
on a relationship between [Fe/H] and Str\"omgren colours
\citep[e.g.][]{Schuster_and_Nissen_1989}. Finally, the sample defined in
\citet{Bovy_etal_2012} is based on $\sim$300 G dwarfs from the Sloan Extension
for Galactic Understanding and Exploration (SEGUE) survey. It should be noted
that the SEGUE survey covers a wider range of galactic radii and significantly
higher galactic scale heights with respect to the other two samples, that are
limited to the solar neighbourhood. This likely explains the larger number of
low-[Fe/H] stars in the Bovy et al. sample. For this sample, the quoted typical
uncertainties are $\sim$0.2~dex for the spectroscopic values of [Fe/H], and
$\sim$0.1~dex for [$\alpha$/Fe].

\begin{figure}
\bc
\resizebox{8.5cm}{!}{\includegraphics{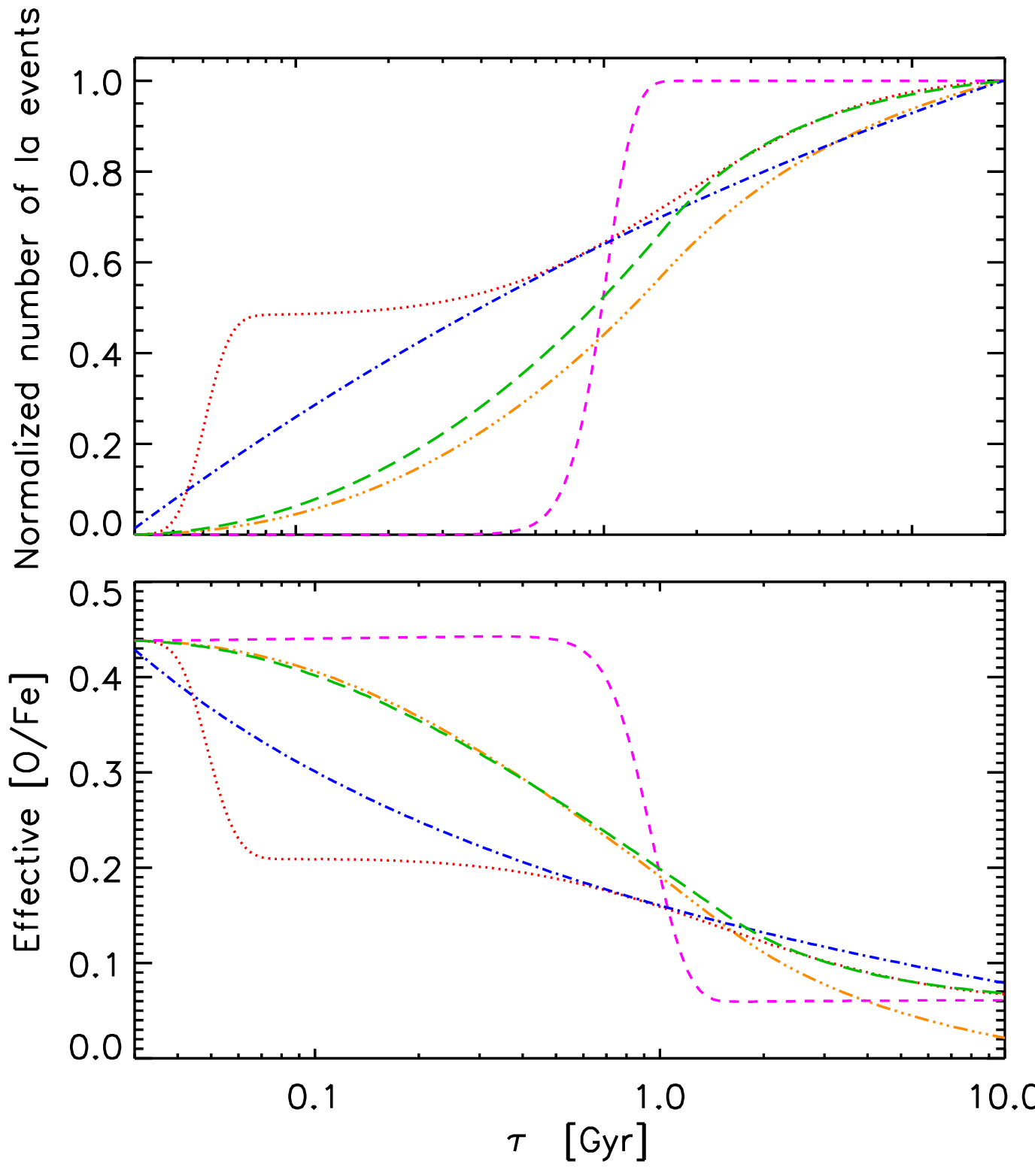}} 
\caption{Normalised number of \SNIa~events (top panel) and effective [O/Fe]
yield (bottom panel) for a simple stellar population at fixed
metallicity. Lines of different colour and style correspond to different
DTDs (the same colour and style coding as in Figs.\ref{fig:dtds}
and \ref{fig:diskmetdistrDTDs} has been used: the blue dot-dashed line
corresponds to the power-law DTD, the magenta dashed line to the narrow
Gaussian DTD, the red dotted line to the bi-modal DTD, the green long dashed
and orange dot-dot-dashed lines to two different variations of the DTD
corresponding to a single degenerate scenario.)}
\label{fig:DTDeffect}
\ec
\end{figure}

For all DTDs used in this study, the peaks of the predicted [Fe/H]
distributions are approximately at the right location (shifted to slightly
higher metallicities with respect to the observational measurements considered
here). The predicted distributions are slightly narrower than the observed
ones, and there is a lack of low-metallicity stars with respect to the sample
by \citet{Wyse_and_Gilmore_1995} and \citet{Bovy_etal_2012}.  One might expect
that the distribution of the entire disk (that is what we predict) is broader
than that of the solar neighbourhood (that is observed). On the other hand, one
should consider that our model distributions are not convolved with the typical
observational uncertainties ($0.1-0.2$~dex, or larger in case of estimates
based on colours), which would tend to broaden them.  In the right panels of
Fig.~\ref{fig:diskmetdistrDTDs}, we compare the observational [$\alpha$/Fe]
distribution by \citet{Bovy_etal_2012} with the model predicted [O/Fe]
distributions (O is the most abundant $\alpha$ element).  In this case, model
predictions vary significantly when changing the adopted DTD. The models that
use the bi-modal and power-law DTDs (red dotted and blue dot-dashed
histograms, corresponding to the red dotted and blue dot-dashed
lines in Fig.~\ref{fig:dtds}) provide narrow distributions peaked around
[O/Fe]$\sim$0.12. In contrast, when adopting a Gaussian DTD (magenta) or a DTD
corresponding to the single degenerate scenario (green long-dashed and
orange dot-dot-dashed), the predicted [O/Fe] distributions are broader,
with a peak at low-[O/Fe] values close to that observed, and a more extended
tail towards higher [O/Fe] values. Our results are consistent with those found
by
\citet{Yates_etal_2013} who included in their analysis three of the DTDs used 
here (the bi-modal, power-law, and Gaussian). As they noted, the lack of any 
prompt component is in contradiction with recent observations 
\citep*[e.g.][]{Maoz_Mannucci_Brandt_2012}. In the following, we will adopt as 
our reference model the one constructed using the narrower single degenerate
DTD shown in Fig.~\ref{fig:dtds} (green long-dashed line). This provides
the best agreement with observational measurements for both distributions shown
in Fig.~\ref{fig:diskmetdistrDTDs}.

The differences between the distributions shown in the right panels of
Fig.~\ref{fig:diskmetdistrDTDs} can be understood by analysing the evolution of
the normalised number of \SNIa~events and of the effective [O/Fe] yield for a
simple stellar population of fixed metallicity. These are shown in
Fig.~\ref{fig:DTDeffect} for different DTDs, using the same colour and
style coding as in Figs.~\ref{fig:dtds} and \ref{fig:diskmetdistrDTDs}. The
abundance distribution results from the convolution of the galaxy's star
formation history with the chemical evolution history, so that a detailed
explanation of the predicted [O/Fe] distribution is not trivial. However, here
we are interested in clarifying qualitatively the origin of the diverse
distributions in the right panels of Fig.~\ref{fig:diskmetdistrDTDs}, so we
discuss the influence of the different DTDs adopted in this study neglecting
the cumulative effect of subsequent stellar generations in order to highlight
the underlying trend.

In the case of a bi-modal DTD (red dotted lines), the contribution from
prompt \SNIa~is significant: the top panel shows that almost half of
the \SNIa~events explode within $\sim$30~Myr. During this time-interval, the
effective [O/Fe] is high, because of the large oxygen contribution
from \SNII. Later on, the effective [O/Fe] yield decreases. It remains
approximately constant for a relatively long time-scale, and then decreases
slowly at late times.  Therefore, one expects a large number of stars with
intermediate/low [O/Fe] and a decreasing fraction at higher values, which is
what we see in the top right panel of Fig.~\ref{fig:diskmetdistrDTDs}. For a
power-law DTD, the behaviour on timescales $t\lesssim 4\times 10^8{\rm yr}$ is
similar to that just described, while the evolution is significantly different
for the other three DTDs considered, that have a significantly lower
contribution from prompt \SNIa. In particular, for a Gaussian DTD, all \SNIa
~events explode after more than $\sim$1~Gyr, in a very short time-interval. The
effective [O/Fe] yield is therefore very high over this time-scale as
significant amount of oxygen are incorporated into the cold gas component
before it starts being enriched with iron by \SNIa. This leads to a prominent
tail at high [O/Fe] values, in addition to a large peak at low values of
[O/Fe]. The behaviour of the two DTDs considered for the single degenerate
model is intermediate, and the smooth evolution visible in
Fig.~\ref{fig:DTDeffect} explains well the continuous distributions shown in
Fig.~\ref{fig:diskmetdistrDTDs}.

\subsection{Yields and time-binning}

\begin{figure*}
\bc
\resizebox{18.cm}{!}{\includegraphics{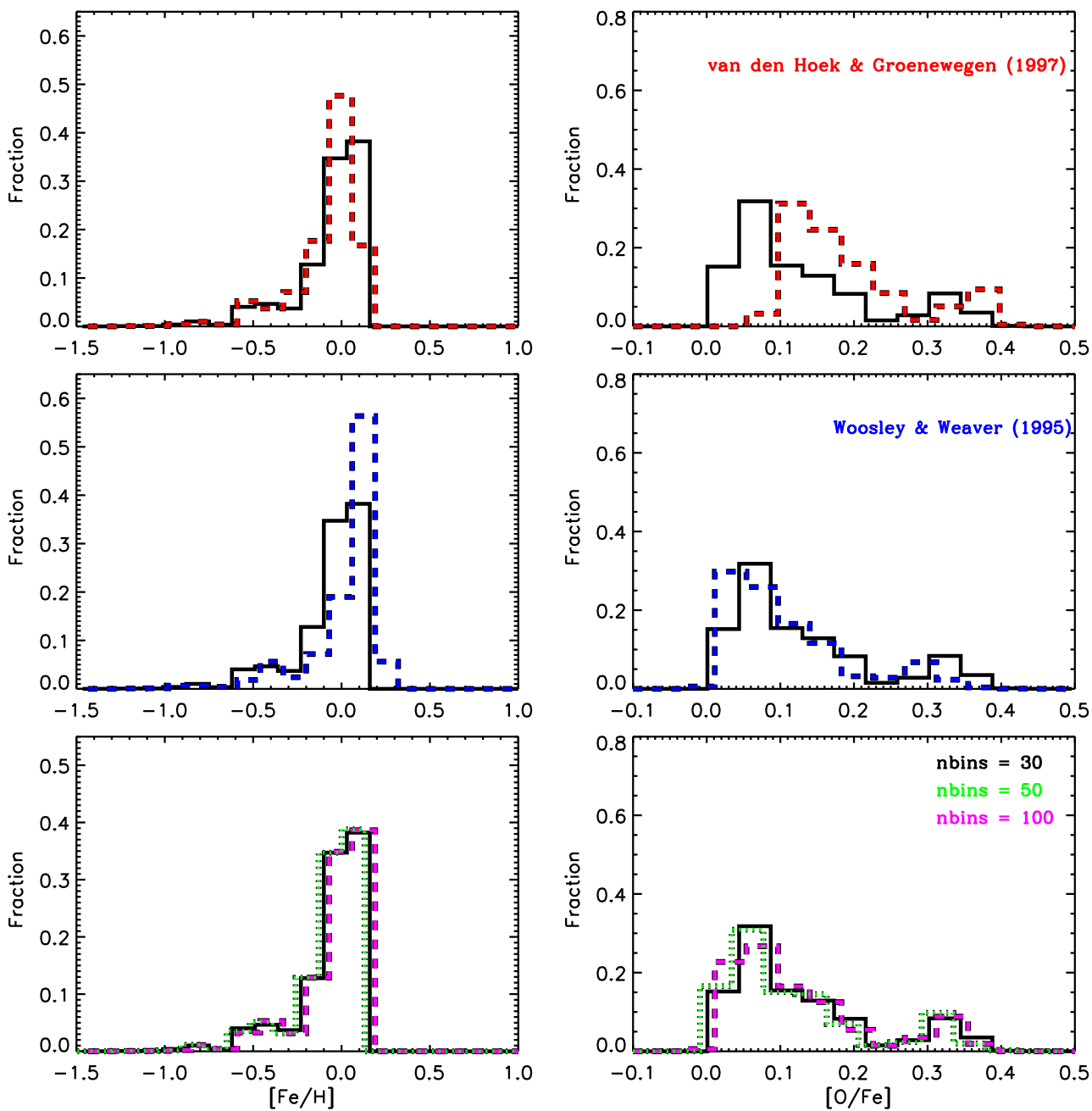}} 
\caption{As in Fig.~\ref{fig:diskmetdistrDTDs}, but this time the black 
  solid histograms refer to our reference model (see text for detail) and
  coloured dashed and dotted histograms correspond to different choices
  for the AGB yields (top panels), \SNII~yields (middle panels), and number of
  time bins for chemical evolution (bottom panels - gren dotted histograms
  show results for 50 timebins and dashed magenta histograms correspond to 100
  timebins). Coloured histograms have been shifted horizontally by 0.03 (-0.03
  for the green dotted line in the bottom panels) in the left panel and
  0.01 (-0.01 for the green dotted line in the bottom panels) in the
  right panel for clarity.}
\label{fig:diskmetdistrYTB}
\ec
\end{figure*}

Fig.~\ref{fig:diskmetdistrYTB} shows the same distributions as in
Fig.~\ref{fig:diskmetdistrDTDs}, but this time the black solid histograms
correspond to the reference model (the green long-dashed line in
Fig.~\ref{fig:dtds}), while the coloured dashed and dotted histograms
show the corresponding distributions obtained by either changing the yields
(those from AGBs in the top panel and those from \SNII~in the middle panel) or
by increasing the number of chemical time bins.

At metallicity around solar (the peak of the [Fe/H] distribution), the
effective [O/Fe] yield by \citet{vandenHoek_and_Groenewegen_1997} is higher
than that adopted in our reference model (that uses the yields
by \citealt{Karakas_2010}). In addition, at variance with the yields
by \citet{vandenHoek_and_Groenewegen_1997}, those by
\citet{Karakas_2010} account for some Fe contribution from \AGB. This explains 
why the [Fe/H] distribution shown by the red dashed histogram in
Fig.~\ref{fig:diskmetdistrYTB} is shifted to slightly lower values, and there
is a lower contribution of low-[O/Fe] stars with respect to the reference
run. The differences between the black solid and blue dashed
histograms in the middle panel of Fig.~\ref{fig:diskmetdistrYTB} is due to
similar reasons: the yields by \citet{Woosley_and_Weaver_1995} give more iron
than the corresponding yields adopted in our reference model
\citep{Chieffi_and_Limongi_2004}, and a slightly lower [O/Fe] effective yield 
at solar metallicity. 

The bottom panels of Fig.~\ref{fig:diskmetdistrYTB} show how the distributions
considered vary increasing the number of chemical time-bins considered. The
three distributions of [Fe/H] almost perfectly overlap (for clarity, we have
introduced a horizontal shift in the figures), while larger variations are
found for the distribution of [O/Fe]. These are due to small inaccuracies in
the timings of re-incorporation of the metals because of the projection on a
discrete array. The bottom panels of Fig.~\ref{fig:diskmetdistrYTB} shows,
however, that our results are stable against a significant increase in the
number of chemical time bins adopted. We have also verified that results do not
change when using alternative choices for the chemical bins (e.g. log-spaced
bins, or fixed `by hand').

\section{Effect of relaxing the instantaneous recycling approximation}
\label{sec:compareIRA}

In the previous section, we have used the lowest resolution simulation
considered in this study to investigate how our model results are influenced by
several ingredients, and have used these results to identify our reference
chemical model. We recall that this adopts the narrower single degenerate DTD
shown in Fig.~1 (the green line) and yields by \citet{Karakas_2010} for low-
and intermediate-mass stars,
\citet{Thielemann_etal_2003} for supernovae Ia; and 
\citet{Chieffi_and_Limongi_2004} for supernovae II. As for our galaxy formation
models, the parameters were kept fixed to those adopted in 
\citet{DeLucia_and_Helmi_2008} and \citet{Li_DeLucia_Helmi_2010}. We will now 
discuss in more detail what are the predictions of our updated chemical model 
for the global properties of Milky Way galaxies, in the higher resolution runs 
used in our study. 

\begin{figure*}
\bc
\resizebox{18.cm}{!}{\includegraphics{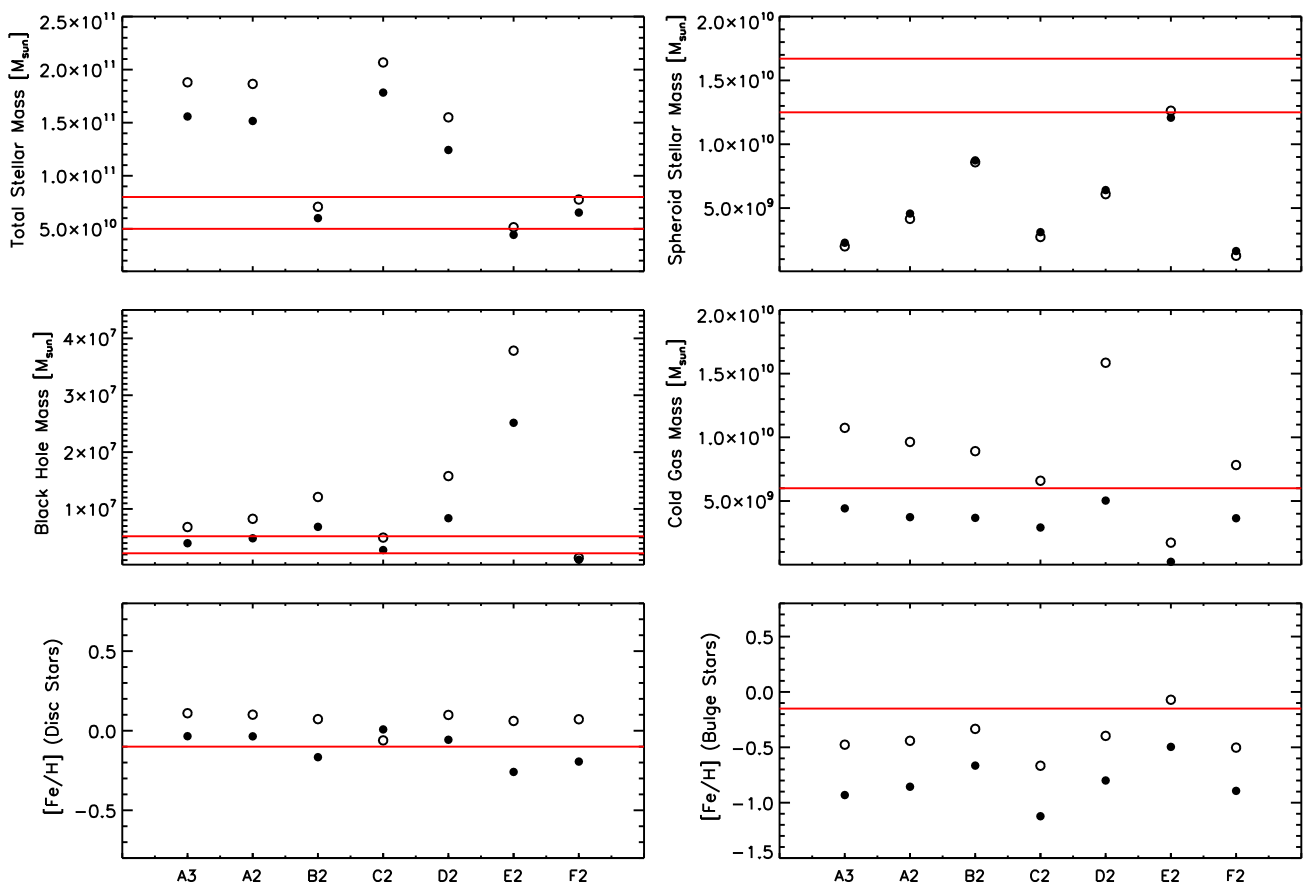}} 
\caption{Physical properties of our model Milky Way galaxies, for the different
  simulations used in this study. Empty circles show results from the model 
  based on the instantaneous recycling approximation, while filled symbols show
  the corresponding results based on the updated chemical model presented in 
  this paper. Red horizontal lines in each panel indicate observational 
  estimates.}
\label{fig:MWprop}
\ec
\end{figure*}

Fig.~\ref{fig:MWprop} shows different physical properties of our model
Milky Way galaxies, for all seven runs considered. Empty circles show the
results from the reference model using the instantaneous recycling
approximation while filled symbols show the corresponding results based on the
updated chemical model presented in this paper. Red horizontal lines in each
panel indicate the observational range/estimates. 

The estimated stellar mass of our Milky Way is $\sim$5$-$8$\times10^{10}\,{\rm
  M}_{\sun}$ \citep[e.g.][]{Flynn_etal_2006}. The top left panel of
  Fig.~\ref{fig:MWprop} shows that only three of the simulations used in this
  study predict stellar masses within this range (these are Aq-B-2, Aq-E-2, and
  Aq-F-2). For all the other runs, the predicted stellar mass is larger than
  the observational estimate by a factor $\sim$2.5.  All runs predict a
  spheroidal stellar mass that is lower than the observational estimate
  ($\sim$25 per cent of the disc stellar mass,
\citealt*{Bissantz_Debattista_Gerhard_2004}), except for the run Aq-E-2 (this 
confirms results found in \citealt{DeLucia_and_Helmi_2008} based on lower 
resolution simulations). We recall that, as discussed in Sec.~\ref{sec:simsam},
our model does not include any prescription for bulge formation during disc 
instability events. This would be required for an appropriate comparison with 
the bulge of our Milky Way. 

The middle left panel of Fig.~\ref{fig:MWprop} shows that the predicted black
hole mass is not far off the observational estimate \citep{Shoedel_etal_2002}, 
but for the run Aq-E-2. This is a consequence of the fact that the galaxy 
formation model has been tuned to reproduce the locally observed relation 
between the mass of the black hole and the mass of the bulge, and that the 
black hole at the centre of our Galaxy is known to be offset with respect to 
such a relation. The amount of cold gas predicted by our model is slightly 
lower than estimated values 
\citep[][and references therein]{Blitz_1997}. Finally, the bottom panels of 
Fig.~\ref{fig:MWprop} show that the predicted average metallicities are not 
far from the observational estimate for the stellar disc but significantly 
lower than suggested by data for the bulge component 
\citep{Freeman_and_Bland-Hawthorn_2002}. We note that the [Fe/H] values 
plotted in these panels for the run adopting an instantaneous recycling 
approximation have been obtained by scaling the total metallicities by the 
appropriate ratio computed using the corresponding run based on our updated 
chemical model. In the following section, we will analyse the metallicity 
distributions predicted for both stellar components in more detail, and 
compare them with observational estimates.

For all runs, our updated chemical model provides stellar masses that are lower
than the corresponding values obtained using the same model parameters in a
scheme that adopts the instantaneous recycling approximation. It is worth
noting, however, that the predicted stellar mass varies significantly depending
on the halo merger history. Also the black hole mass, the gas mass and the mean
metallicity of both the disc and bulge stellar components are systematically
lower in the updated chemical model, while the stellar masses of the spheroidal
component are slightly larger in this model with respect to the corresponding
values obtained by adopting the instantaneous recycling approximation. We will
come back to the difference between the two schemes at the end of this section.

\begin{figure*}
\bc
\resizebox{18.cm}{!}{\includegraphics{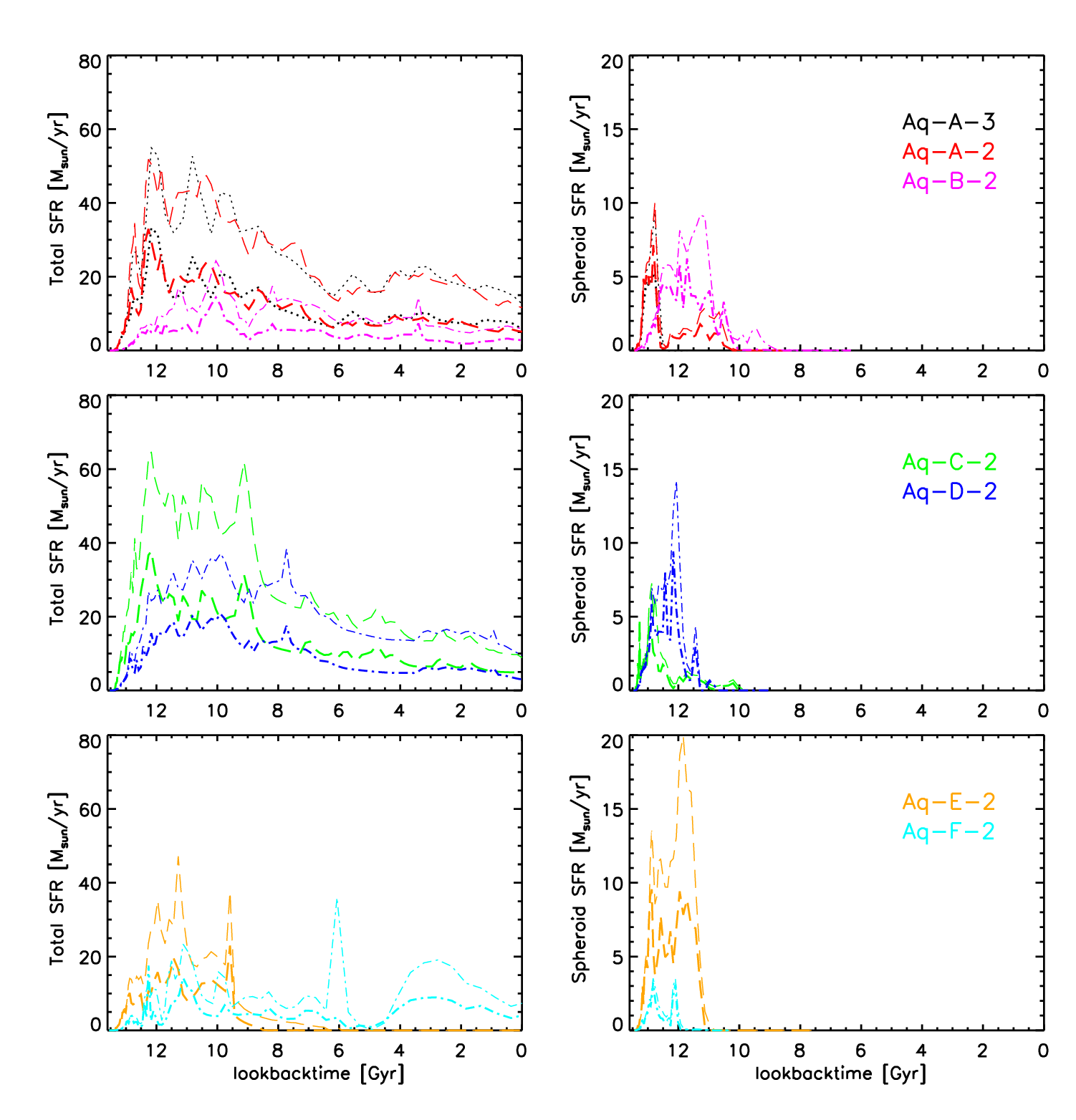}} 
\caption{Star formation history of our model Milky Way galaxies, from the
  different simulations used in this study. The left panel shows the total star
  formation history, while the right panel corresponds to the star formation
  history of the spheroidal component. Thin lines show results from the
  model based on the instantaneous recycling approximation, while thick
  lines show the corresponding results based on the updated chemical model
  presented in this paper. Dotted lines are used for Aq-A-3, while for the
  level 2 simulations, dashed and dot-dashed lines are used for the first and
  second simulation indicated in the legend of each row.}
\label{fig:SFR}
\ec
\end{figure*}

Fig.~\ref{fig:SFR} shows the total star formation history of our model Milky
Way galaxies (this has been obtained by summing, at each time, the star
formation rates of all progenitors of the final Milky Way galaxy, left panels),
and that of the stars in the spheroidal components (right panels), for all
seven runs used in this study.  Results from the two resolution levels
considered for run Aq-A show very good convergence (compare black dotted
and red dashed lines in the top panels).  For all runs considered in this
study, the total star formation history is fairly extended, with present day
values ranging between almost zero for Aq-E-2 and $\sim$6$\,{\rm
M}_{\sun}\,{\rm yr}^{-1}$ for Aq-A-2. With the exception of Aq-E-2, the present
level of star formation activity is still somewhat higher than the
$\sim$2$\,{\rm M}_{\sun}\,{\rm yr}^{-1}$ estimated for the mean star formation
rate in the Milky Way disc
\citep{Chomiuk_and_Povich_2011}, but systematically lower than the 
corresponding values found in our previous model based on the instantaneous
recycling approximation. All stars that currently reside in the spheroidal
component of our model Milky Way galaxies formed more than $\sim$10~Gyr ago.
The peak of the star formation is never higher than $\sim$10$\,{\rm
M}_{\sun}\,{\rm yr}^{-1}$, while the typical width ranges between 1 and 3
Gyr. As we will see in the next section, this has important consequences for
the metallicity distribution of the spheroidal component.  Generally, the star
formation rates from our updated chemical model are systematically lower than
those obtained from the corresponding runs assuming an instantaneous recycling
approximation, at all times. For the spheroidal component, this systematic
difference is lower, and negligible for the runs Aq-A-2, Aq-C-2, and Aq-F-2.

\begin{figure}
\bc
\resizebox{8.5cm}{!}{\includegraphics{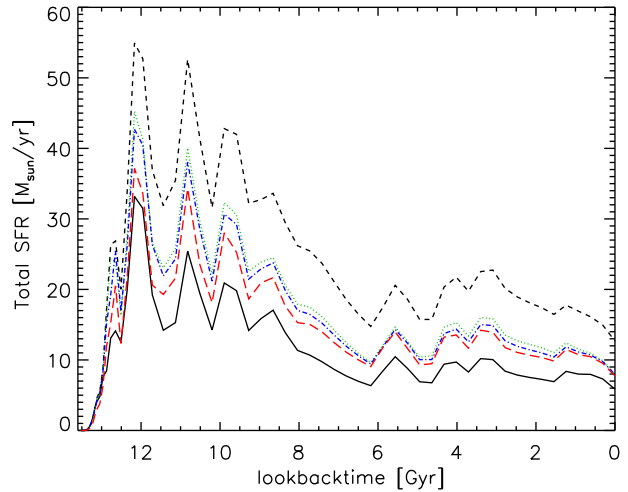}} 
\caption{Star formation history for the model Milky Way galaxy in the run 
Aq-A-3. The solid black line shows the prediction from the reference model used
in this study, while the black dashed line shows the corresponding prediction
from a run that adopts an instantaneous recycling approximation. The coloured
dotted, dot-dashed and long-dashed lines corresponds to variations of the
latter run with varying parameters (see text for details).}
\label{fig:SFRAq3}
\ec
\end{figure}

Naively, the lower star formation rate in the updated chemical scheme can be
attributed to the slower recycling of gas from massive stars with respect to a
scheme adopting an instantaneous recycling approximation. To better understand
the differences between the two schemes, and quantify them in terms of (i)
amount of cold gas recycled from stars, (ii) chemical yields, and (iii) energy
injected into the cold gas by supernovae explosions, we have carried out
several additional runs using Aq-A-3, and varying the relevant
parameters in the run assuming an instantaneous recycling approximation. As 
explained in Section~\ref{sec:simsam}, this model assumes an instantaneous 
recycling fraction $R=0.43$ (corresponding to the integral contribution of 
H$+$He from \SNII, \SNIa, and \AGB~for the Chabrier IMF assumed in our 
study), a constant yield $Y=0.03$, and a number of supernova events per unit 
stellar mass formed $\eta_{\rm SN} = 8\times10^{-3}$. 

The amount $R$ of cold gas restored by supernova explosions has a major effect 
since it directly and promptly affects the reservoir of gas available for star 
formation. The effective yield $Y$ acts on the star formation through the 
dependence on the metallicity of the cooling function, so that the higher 
its value the shorter is the cooling time. Finally, the energy injected by 
supernova explosions delays or prevents star formation in the surrounding gas 
by reheating part of it. 

In order to show the cumulative effect of these quantities, we changed them one
at a time in the run adopting the instantaneous recycling approximation, so as
to be consistent with the values used in the updated chemical model. In this
case, the recycled fraction $R$ is consistent with that adopted in the run with
instantaneous recycling, with a weak dependence on the adopted yields and the
metallicity of the star. However, in the updated chemical scheme, this fraction
of gas is ejected into the insterstellar medium over about 10~Gyr, as discussed
in previous sections. In order to mimic the effect of this dilution in the run
based on the instantaneous recycling approximation, we set $R=0.15$ that is the
fraction of H$+$He restored by \SNII~only. As for the effective yield, our
tables are consistent with the assumption $Y=0.02$. Finally, we assume
$\eta_{\rm SN}=0.014$, that is the total number of both
\SNII~($\eta_{\rm SNII}=0.012$) and \SNIa~($\eta_{\rm SNII}=0.0018$) events per 
unit stellar mass formed.

Results of this analysis are shown in Fig.~\ref{fig:SFRAq3}. The dashed and
solid black lines show the two reference runs with an instantaneous recycling
approximation and updated chemical model, respectively. The dotted green
line in Fig.~\ref{fig:SFRAq3} corresponds to a run adopting the instantaneous
recycling approximation but $R=0.15$. In this run, the star formation rates are
reduced with respect to the dashed black lines at all times, but are still
significantly higher than those corresponding to our updated chemical model.
The dot-dashed blue line shows the results obtained by further reducing
the effective yield to $Y=0.02$, that is closer to that adopted in our chemical
model. Finally, the long-dashed red line corresponds to a run that adopts
$R=0.15$, $Y=0.02$, and $\eta_{\rm SN} = 0.014$ that is the appropriate number
based on our choice of IMF and model for \SNIa. At early times (lookback time
larger $\sim$11.5~Gyr), the star formation rates obtained are very close to
those resulting from our updated chemical model. At late times, the results
tend to diverge because of the contribution from late recycling of gas and
metals, and its effects on the metallicity of the hot gas and gas cooling
rates.

\section{Milky Way disc, spheroid and satellites}
\label{sec:results}

\begin{figure*}
\bc
\resizebox{18.cm}{!}{\includegraphics{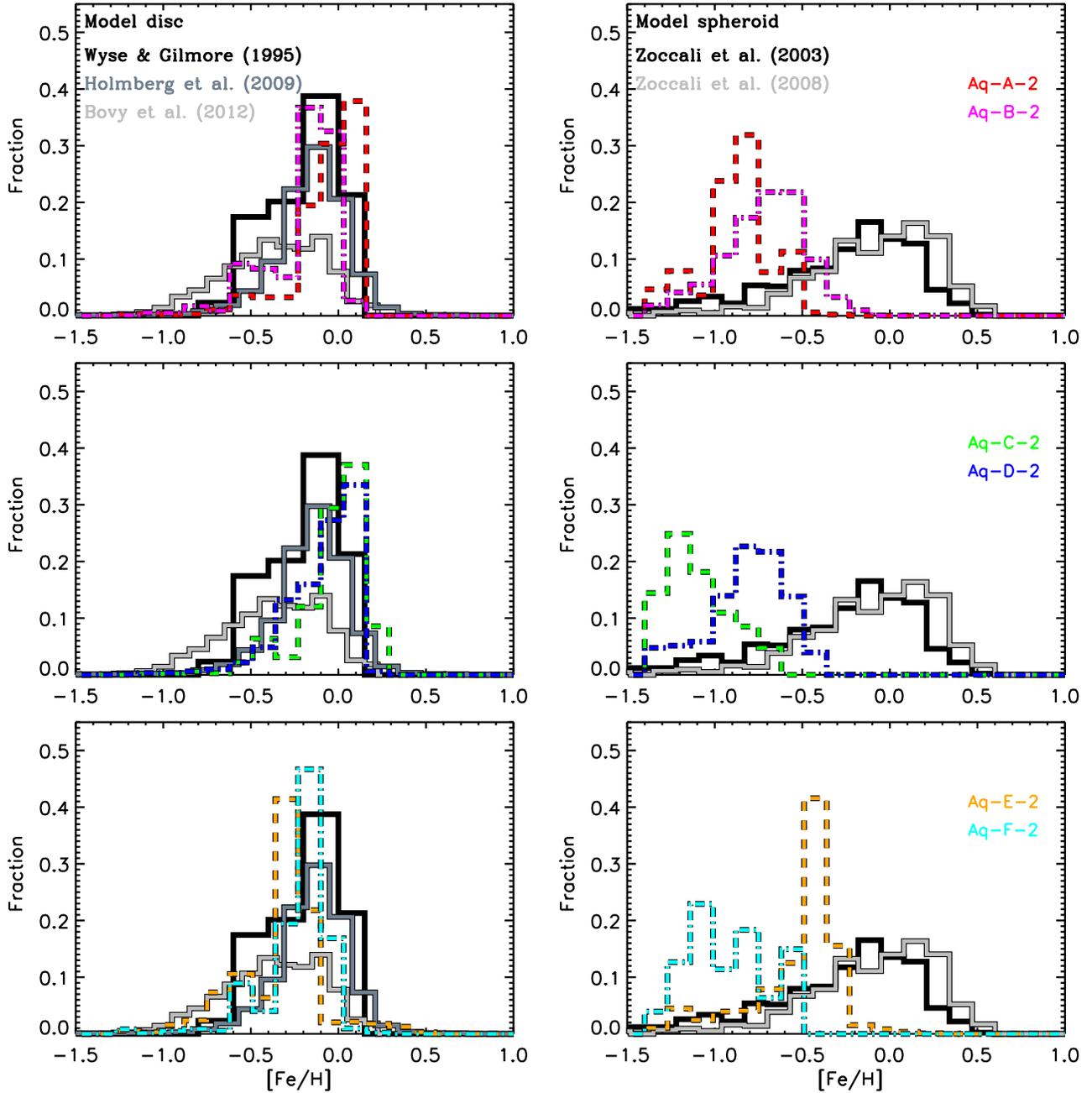}} 
\caption{[Fe/H] distributions of the stars in the stellar disc (left panels) 
  and spheroid (right panel) for all the simulations used in this study (
  dashed and dot-dashed histograms corresponding to the first and second
  simulation indicated in the legend for each row, respectively), compared with
  different observational determinations (black and gray histograms).}
\label{fig:diskbulgeFeH}
\ec
\end{figure*}

In the previous sections, we have analysed in detail the dependence of our
results on various model ingredients, as well as the global properties and star
formation histories of all runs used in our study. We now turn to the predicted
metallicity distributions of the disc and spheroid components. The [Fe/H]
distributions are shown in Fig.~\ref{fig:diskbulgeFeH}. Left panels correspond
to the model disc, while right panels to the spheroid. Black and grey
histograms show different observational measurements, and coloured dashed
and dot-dashed lines show model predictions for the six level-2 resolution
haloes considered in this study.

We remind that we have `tuned' our model by considering primarily the [Fe/H]
and [O/Fe] metallicity distributions of the disc component for the run
Aq-A-3. The left panels of Fig.~\ref{fig:diskbulgeFeH} show that, for the same
combination of model parameters and ingredients, similar [Fe/H] distributions
are obtained for all six haloes considered. Considering the uncertainties in
the observational data, all of them appear to be in reasonable agreement with
measurements of our Milky Way disc stars. This is not surprising given the
similar behaviour of star formation histories plotted in the left panels of
Fig.~\ref{fig:SFR}. Aq-A-2 and Aq-C-2 have very similar star formation
histories, with a slightly more significant peak at early cosmic epochs for the
run Aq-A-2. This explains why the red and green dashed histograms (in the top
and middle panels) in the left panels of Fig.~\ref{fig:diskbulgeFeH} are very
similar with a slightly larger fraction of high [Fe/H] stars in the run
Aq-C-2. For the run Aq-D-2, the star formation history is broader, and so is
the corresponding [Fe/H] distribution. Aq-B-2 is again very similar to Aq-A-2
but the overall level of star formation rate is lower, which explains why the
peak of the [Fe/H] distribution is shifted to slightly lower values of [Fe/H]
with respect to that of Aq-A-2. Finally, for the runs Aq-E-2 and Aq-F-2, the
star formation rates are always lower than $\sim$20$\,{\rm M}_{\sun}\,{\rm
yr}^{-1}$. The star formation rate drops to virtually zero after the first
$\sim$4~Gyr for the run Aq-E-2, and shows a late `bump' of star formation for
the run Aq-F-2. As a consequence, the [Fe/H] distributions are very narrow,
with a somewhat larger contribution from stars with high [Fe/H] in the run
Aq-F-2.

For all six level-2 haloes, the star formation history of the spheroid
component is relatively narrow and peaked at high redshift. In addition, the
actual values of the star formation rate are relatively modest. This translates
into metallicity distributions that are offset low with respect to the
observational estimates. The largest fraction of high-[Fe/H] stars is obtained
for the run Aq-E-2 whose spheroid component exhibits a relatively broad
(compared to the other runs) star formation history at levels that are the
highest compared to those of the other runs. For this run, however, the peak of
the [Fe/H] metallicity distribution is also lower than the observed location by
about $0.5$~dex. As noted in the previous section, our model neglects any
contribution to bulge formation through disc instabilities. If these occur late
during the evolution of the Milky Way, they could contribute to populating the
high-[Fe/H] tail of the distributions in our model. Given the uncertainties in
modelling dynamical disc instabilities, and the fine-tuning that appears to be
required to reproduce the observational measurements, we prefer not to
investigate this issue further in this work.

\begin{figure}
\bc
\resizebox{8.5cm}{!}{\includegraphics{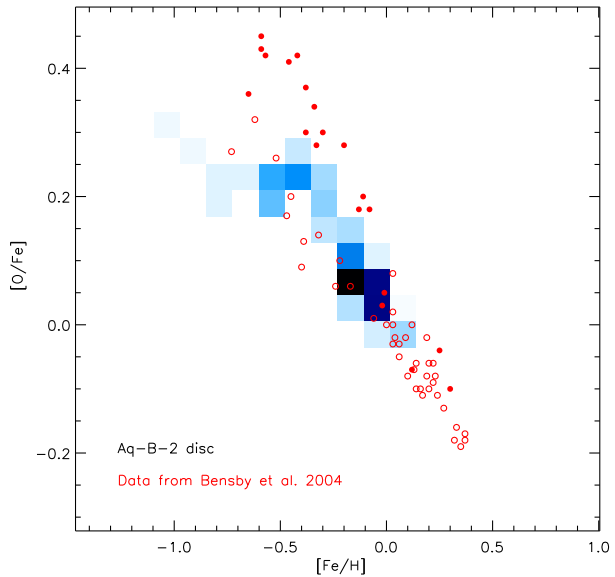}} 
\caption{[O/Fe] versus [Fe/H] distribution for the stars in the stellar disc of
  the simulation Aq-B-2. Red circles show observational measurements
  from \citet*{Bensby_Feltzing_Lundstrom_2004}, with filled and open symbols
  corresponding to thick and thin disc stars, respectively.}
\label{fig:B2diskOFeFeH}
\ec
\end{figure}

\begin{figure}
\bc
\resizebox{8.5cm}{!}{\includegraphics{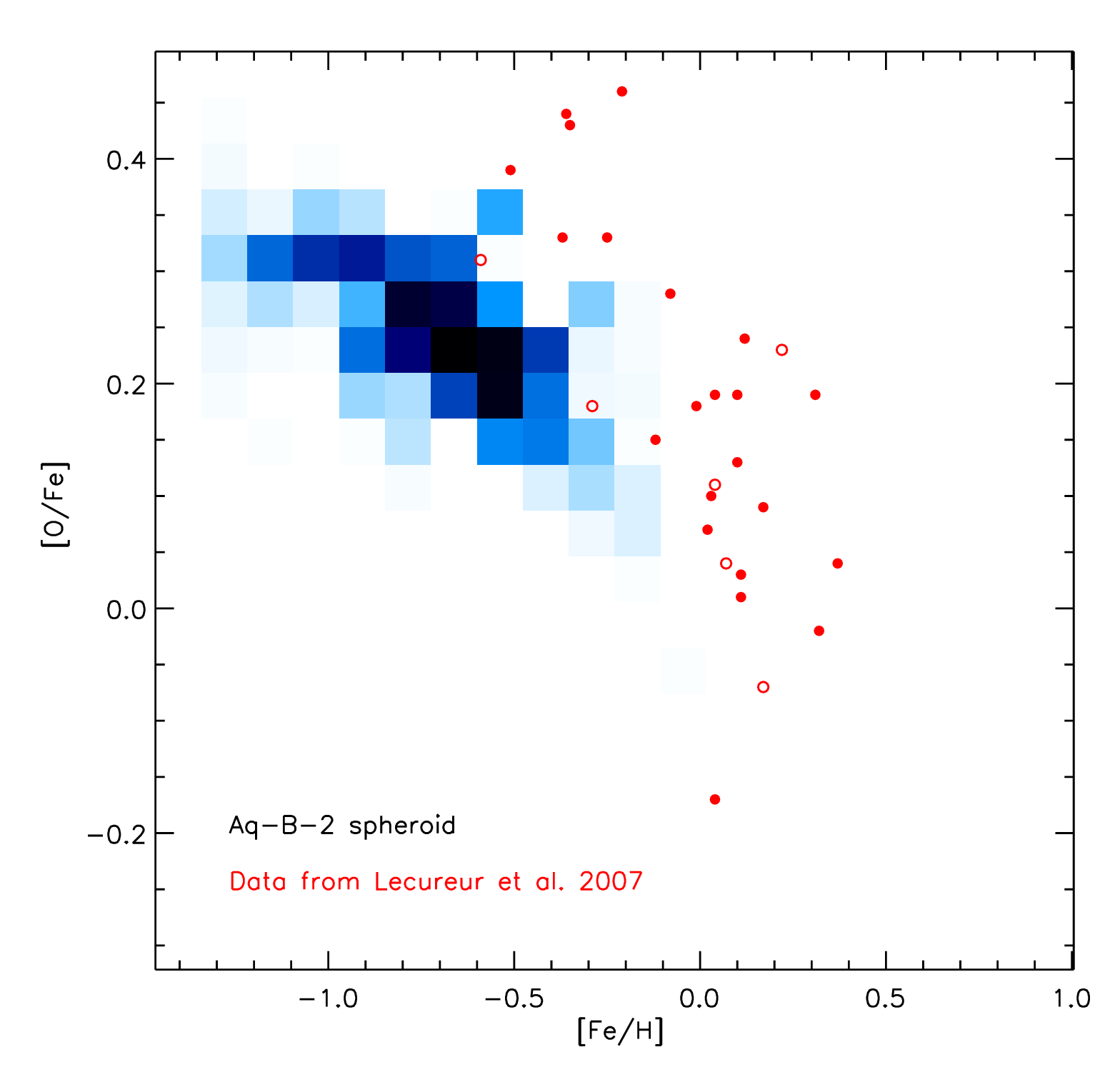}} 
\caption{As for Fig.~\ref{fig:B2diskOFeFeH} but for the stars in the spheroid
  of the simulation Aq-B-2. Red circles show observational measurements
  from \citet{Lecureur_etal_2007}, with empty symbols corresponding to
  uncertain O abundances.}
\label{fig:B2bulgeOFeFeH}
\ec
\end{figure}

Do any of the runs considered in this study give a model galaxy that is in
reasonable agreement with both the global properties and the metallicity
distributions of our Milky Way? Forgetting about the too low metallicity of the
spheroid component, the run that provides the best match is probably Aq-B-2. For
this run, the predicted stellar mass of the model Milky Way galaxy is 
$\sim$6$\times 10^{10}\,{\rm M}_{\sun}$, in very good agreement with the 
estimated stellar mass for our Galaxy. The mass of the spheroidal component 
(we recall that our runs do not include any disc instability channel) is 
$\sim$8.7$\times 10^{9}\,{\rm M}_{\sun}$, slightly below the observationally 
estimated value. The current level of star formation rate is $\sim$2.8$\,{\rm
M}_{\sun}\,{\rm yr}^{-1}$, and the present value of the
\SNIa~rate is $0.58$ events/century (two or three times larger than the  
observational estimate of $\sim$0.2$-$0.3 events/century - see e.g. 
\citealt*{Cappellaro_Evans_Turatto_1999} and \citealt{Li_etal_2011}). 
The recent work by \citet{Starkenburg_etal_2013}, based on our reference model
with the instantaneous recycling approximation, reached the same conclusion as
for the galaxy that most closely resembles the Milky Way. Interestingly, they
showed that the same run has also a satellite luminosity function in quite good
agreement with observational measurements but does not contain satellites as
bright as the Small or Large Magellanic Clouds. As they note (see their Section
6.2), with the adopted star formation and feedback prescriptions, our best
Milky Way galaxy corresponds to a dark matter halo with mass $8\times
10^{11}\,{\rm M}_{\sun}$, that is on the low end of observational
estimates \citep{Smith_etal_2007,Xue_etal_2008,Li_and_White_2008}. In addition,
over the mass range probed by the Aquarius haloes, the relation between dark
matter halo mass and stellar mass for central galaxies is offset with respect
to that derived using abundance matching methods by e.g. \citet{Guo_etal_2010}
that is reproduced (by construction) in the model by \citet{Guo_etal_2011}.

Figs.~\ref{fig:B2diskOFeFeH} and \ref{fig:B2bulgeOFeFeH} show the [O/Fe] versus
[Fe/H] metallicity distributions for both the stellar disc and the spheroid
component of the model Milky Way galaxy in the run Aq-B-2. The 2D distributions
from the other runs are actually very similar. As discussed above, the [Fe/H]
metallicity distribution is in quite good agreement with observational data of
the thin disc (in part by construction). In contrast, stars in the spheroid are
significantly metal poorer than observed. Finally, stars in both components
cover approximately the same range of [O/Fe] abundances as the data.

\begin{figure}
\bc
\hspace{-0.5cm}
\resizebox{8.5cm}{!}{\includegraphics{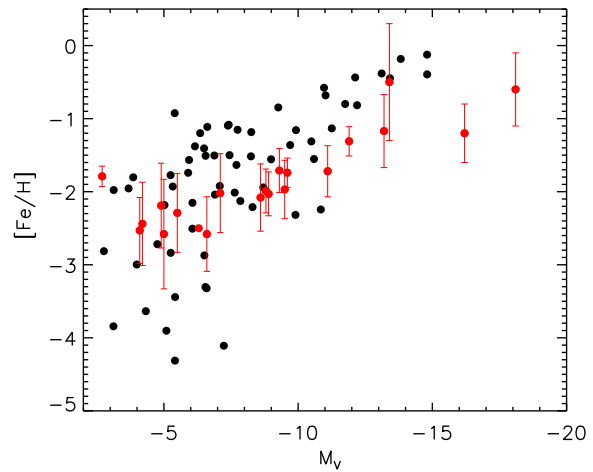}} 
\caption{Metallicity-luminosity relation for the satellites. Filled black 
symbols show model predictions for all satellite galaxies in the Aq-B-2 run
within 280 kpc from the central model Milky Way. Red filled symbols with error
bar show observational measurements for both the `classical' and ultra-faint
satellites. Model data points have been shifted by -0.23 dex (see text for 
details).}
\label{fig:metlumsat}
\ec
\end{figure}

Fig.~\ref{fig:metlumsat} shows the metallicity-luminosity relation for all the 
satellite galaxies of the Aq-B-2 run within 280 kpc from the central galaxy. 
Filled black circles show model predictions while red filled symbols with error
bars show observational measurements for both the `classical' and ultra-faint
satellites (we have used the compilation of literature data by 
\citealt{Li_DeLucia_Helmi_2010}). As noted by \citet{Starkenburg_etal_2013}, 
our metallicities are mass-weighted while the observational measurements are
mean [Fe/H], that they find to be on average 0.23 dex lower than the logarithm
of the average over the ratio of Fe to H. Therefore, we have shifted by
$-0.23$~dex the model data points in Fig.~\ref{fig:metlumsat} (this shift,
however, is not significant on the scale shown).

As mentioned above, the run Aq-B-2 does not contain satellites as bright as the
two brightest members of the Local Group (the LMC and SMC). Overall, the model
metallicity-luminosity relation is in relatively good agreement with data,
although the model relation appears to be offset slightly higher and to have a
somewhat steeper slope than observed. Nevertheless, the level of agreement with
data shown in Fig.~\ref{fig:metlumsat} is remarkable, given that these data
were not considered when tuning our chemical model. It remains to be seen if
the same model is able to reproduce also the vast amount of chemical data
available for the more general galaxy population, both in the local Universe
and at higher redshift. We will address this issue in future work.

\section{Discussion and Conclusions}
\label{sec:discconcl}
 
We have presented a new method to account for relaxing the instantaneous 
recycling approximation and trace individual elemental abundances within a
semi-analytic model of galaxy formation and evolution. The model we use in 
this paper is based on the model described in detail in 
\citet{DeLucia_and_Blaizot_2007} but has been refined to follow more accurately
processes on the scale of the Milky Way's satellites (for details, we
refer to \citealt{DeLucia_and_Helmi_2008}, \citealt{Li_DeLucia_Helmi_2010}, and
\citealt{Starkenburg_etal_2013}). We take advantage of the simulations carried 
out within the Aquarius Project \citep{Springel_etal_2008}: a sample of six
haloes of roughly Milky Way mass and with no massive close neighbour at $z=0$,
all simulated at the same level of resolution (corresponding to a particle mass
between $m_p= 6.45\times10^3$ and $m_p=1.40\times10^4\,{\rm M}_{\sun}$).  To
analyse the convergence of the physical model and study the influence of
different model ingredients, we also take advantage of one lower resolution
run, with a particle mass of $m_p = 4.91\times 10^4\,{\rm M}_{\sun}$.

All previously published models that implement a detailed description of
chemical
evolution \citep{Nagashima_etal_2005,Arrigoni_etal_2010,Benson_2012,Yates_etal_2013}
adopt a similar method to account for the finite lifetime of stars: the past
star formation history is re-binned and stored in memory. This information is
then used to compute the metal restitution rates at any time during model
integration. Since the calculation requires an averaging of individual star
formation events, it becomes increasingly inaccurate as the time elapsed
between star formation events increases, i.e. for long living stars whose
accounting represents the main reason to implement an algorithm that relaxes
the instantaneous recycling approximation. We have developed a new method that
instead computes the metal ejection rates every time new stars are formed, and
stores this information in the future. While a binning is also necessary (to
avoid memory overheads), the method allows an accurate accounting of the
timings and properties of individual star formation events.

The model extension discussed above requires the introduction of new model
ingredients, and a number of assumptions. We have analysed, in particular, the
influence of (i) different distribution time delays (DTDs) for \SNIa; (ii)
different yields; (iii) number of chemical time bins adopted. These have an
important influence on the predicted [O/Fe] distributions for stars in the disc
of the model Milky Way galaxies. In particular, DTDs with a less significant
prompt component predict broader distributions with a more extended tail
towards high [O/Fe] values. This is because significant amounts of oxygen
(produced by
\SNII) can be incorporated into the cold gas component before it starts being
enriched with iron by \SNIa. Our reference chemical model assumes a DTD based
on the single degenerate model described by \citet{Matteucci_and_Recchi_2001}
and corresponds to a fraction of prompt supernovae Type Ia of about 5 per cent
(if `prompt' is defined as exploding within $10^8$~yr from the star formation
episode, the fraction increases to about 23 per cent when considering all 
\SNIa~events within $4\times10^8$~yr). Our findings are consistent with (but 
stronger than) results from \citet{Yates_etal_2013} who require $\leq 50$ 
per cent of the \SNIa~within $4\times10^8$~yr. Different sets of yields for 
\SNII~and \AGB~can affect significantly the metallicity distributions of disc 
stars, depending on the relative contribution of iron and oxygen. We have shown
that our results are stable against a significant increase in the number of
adopted chemical time bins, as well as against the choice of time-bins. In
particular, in our reference model, bins are chosen in such a way that each
contains a constant number of events. Results do not change if bins are
log-spaced in time, or simply fixed `by hand'.

There are degeneracies in the chemical model ingredients: e.g. a different DTD
than the one adopted in our reference model, combined with a different set of
yields might provide very similar results in terms of the metallicity
distribution. However, once a reference model has been chosen, the results
listed below remain valid. In particular, tuning the model considering mainly
the [Fe/H] and [O/Fe] metallicity distributions of the disc component in the
lowest resolution run, we find that:

\begin{itemize}

\item our updated chemical model results in a systematically lower star 
formation history with respect to runs adopting the same physical model but 
based on an instantaneous recycling approximation. This is a consequence of 
a lower effective yield and number of supernovae events in our updated chemical
model, as well as of the slower recycling of gas, metals and energy from 
massive stars. The latter has important consequences on the metallicity of the 
hot gas component and gas cooling rates. 

\item For all runs, the [Fe/H] metallicity distributions of the disc stars are 
in reasonable agreement with observational measurements. The star formation
history of the spheroid component is relatively narrow and peaked at high
redshift.  This results into a metallicity distribution that is offset low with
respect to observational data. Our runs do not include a disc instability
channel for bulge formation. These events could help to populate the high
[Fe/H] tail of the metallicity distributions, but a certain level of
fine-tuning appears to be required in order to reproduce the observational
data.

\item The six haloes used in this study provide a range of predicted 
properties for the model Milky Way. Among these, the one that most resembles 
our Galaxy is Aq-B-2. For this particular run, we have shown that not only the
global properties of the Milky Way are well reproduced, but also the observed
relation between the average metallicity of the satellite galaxies and their
luminosity (albeit with a slightly steeper slope).

\end{itemize}

In this paper, we have only applied our updated chemical scheme to Milky 
Way-like haloes and analysed the dependence on different model ingredients 
and basic predictions. In future work, we plan to extend the analysis to other
components of the Milky Way (in particular the stellar halo), as well as to
different cosmic epochs and galaxy types. By analysing the metallicity 
distributions in different baryonic components, and the dependence on the 
feedback and recycling scheme, our model should be able to provide important
constraints on these physical processes.

\section*{Acknowledgements}
GDL acknowledges financial support from the European Research Council under the
European Community's Seventh Framework Programme (FP7/2007-2013)/ERC grant
agreement n. 202781. We acknowledge fruitful and stimulating discussions with
Francesca Matteucci, Simone Recchi, and Donatella Romano. We thank J. Bovy for
providing us with his data in electronic format, and S. Recchi, D. Romano, R. 
Yates, and E. Starkenburg for constructive comments on a preliminary version of
this manuscript. We are indebted to Volker Springel for making the merger trees
from the Aquarius haloes available to us. We thank the referee, Christopher Few,
for his constructive and positive feedback. 

\bsp

\label{lastpage}

\bibliographystyle{mn2e}
\bibliography{chemistry}

\appendix

\section{Sliding and re-binning of the metal array}

In general, the size of the sub-step $\delta\tau$ and of the \metar timebins
are independent from each other, and generally treated as free parameters that
can be modified according to memory and computational time constraints. In
addition, $\delta\tau$ can change from one snapshot to another if the snapshot
times are not equally spaced. Therefore, in general, $\delta\tau$ could
correspond to 0 or several entries in \metar or will fall within one of its
entries.

When the code evolves the galaxy $G$ from a sub-step to the next one, a
fraction $f = \delta\tau / \Delta T^A_0$ of the relevant entry of the array
\metar is re-incorporated (and subtracted from \metar). Then \metar should be
shifted forward in time by $\delta\tau$, and all its entries should be re-binned
so that a fraction $f_j=\delta\tau / \Delta T^A_j$ of the $j$th entry flows in
the previous entry \metar$[j-1]$. It easy to compute that the first entry
should contain 
\begin{displaymath}
Z_0\times(1-f_0) + Z_1\times f_1
\end{displaymath}
and that, after another $\delta\tau$, it should contain:
\begin{displaymath}
Z_0\times(1-2\times f_0) + Z_1\times f_1 \times (\-f_0) + Z_1\times f_1
\end{displaymath}
Therefore, a correct re-binning requires a separate storage of the $Z_j$
elements at each sub-step, which again leads to a dramatic memory overhead. In
addition, if new stars are formed, new metals are produced and should be added
to the various entries of the \metar array requiring additional information to
be stored for an appropriate re-binning. To avoid this problem, as explained in
Section~\ref{sec:future}, we have simply kept the \metar fixed in cosmic time
between two subsequent snapshots, and re-binned it only when the present time
coincides with the cosmic time corresponding to a snapshot.

\end{document}